\documentclass[11pt,14paper]{article}

\usepackage{jheppub}
\usepackage[T1]{fontenc}
\usepackage{amsfonts,bm}
\usepackage[dvipsnames]{xcolor}
\usepackage[normalem]{ulem}

\usepackage{orcidlink}

\graphicspath{{Figs/}}

\newcommand{\jp}{J/\psi}
\newcommand{\X}{X(3872)}
\makeatletter
\renewcommand*\env@matrix[1][\arraystretch]{
  \edef\arraystretch{#1}
  \hskip -\arraycolsep
  \let\@ifnextchar\new@ifnextchar
  \array{*\c@MaxMatrixCols c}}
\makeatother

\title{
Predicting isovector charmonium-like states from \texorpdfstring{$X(3872)$}{X(3872)} properties
}

\author[a,b]{Zhen-Hua Zhang,\orcidlink{0000-0001-6072-5378}}

\author[c]{Teng Ji,\orcidlink{0000-0003-0366-1042}}

\author[c]{Xiang-Kun Dong,\orcidlink{0000-0001-6392-7143}}

\author[a,b,d,e]{Feng-Kun~Guo,\orcidlink{0000-0002-2919-2064}}

\author[f]{Christoph~Hanhart,\orcidlink{0000-0002-3509-2473}}

\author[c,f,g]{Ulf-G.~Mei{\ss}ner\orcidlink{0000-0003-1254-442X}}

\author[c,g]{and Akaki Rusetsky\orcidlink{0000-0002-4288-8672}}

\affiliation[a]{CAS Key Laboratory of Theoretical Physics, Institute of Theoretical Physics, Chinese Academy of Sciences,\\ Zhong Guan Cun East Street 55, Beijing 100190, China}

\affiliation[b]{School of Physical Sciences, University of Chinese Academy of Sciences, Beijing 100049, China}

\affiliation[c]{Helmholtz-Institut f\"ur Strahlen- und Kernphysik and Bethe Center for Theoretical Physics,\\ Universit\"at Bonn, D-53115 Bonn, Germany}

\affiliation[d]{Peng Huanwu Collaborative Center for Research and Education, Beihang University, Beijing 100191, China}

\affiliation[e]{Southern Center for Nuclear-Science Theory (SCNT), Institute of Modern Physics,\\ Chinese Academy of Sciences, Huizhou 516000, China}

\affiliation[f]{Institute for Advanced Simulation (IAS-4), Forschungszentrum J\"ulich, D-52425 J\"ulich, Germany}

\affiliation[g]{Tbilisi State University, 0186 Tbilisi, Georgia}

\emailAdd{zhangzhenhua@itp.ac.cn}
\emailAdd{teng@hiskp.uni-bonn.de}
\emailAdd{fkguo@itp.ac.cn}
\emailAdd{xiangkun@hiskp.uni-bonn.de}

\abstract{
Using chiral effective field theory, we predict that there must be isovector charmonium-like $D\bar D^*$ hadronic molecules with $J^{PC}=1^{++}$ denoted as $W_{c1}$. The inputs are the properties of the $X(3872)$, including its mass and the ratio of its branching fractions of decays into $J/\psi\rho^0$ and $J/\psi\omega$. 
The predicted states are virtual state poles of the scattering matrix, pointing at a molecular nature of the $X(3872)$ as well as its spin partners. They should show up as either a mild cusp or dip at the $D\bar D^*$ thresholds, explaining why they are elusive in experiments. 
The so far negative observation also indicates that the $X(3872)$ is either
a bound state with non-vanishing binding energy or a virtual state,
only in these cases the $X(3872)$ signal dominates over that from the $W_{c1}^0$.
The pole positions are 
$3881.2^{+0.8}_{-0.0}- i 1.6^{+0.7}_{-0.9}$~MeV for $W_{c1}^0$ on the fourth Riemann sheet of the $D^0\bar D^{*0}$-$D^+D^{*-}$ coupled-channel system, and  $3866.9^{+4.6}_{-7.7}- i (0.07\pm0.01)$~MeV for $W_{c1}^\pm$ on the second Riemann sheet of the $(D\bar D^*)^\pm$ single-channel system.
The findings imply that the peak in the  $J/\psi\pi^+\pi^-$ invariant mass distribution is not purely from the $X(3872)$ but contains contributions from $W_{c1}^0$ predicted here.
The states should have isovector heavy quark spin partners with $J^{PC}=0^{++}$, $2^{++}$ and $1^{+-}$, with the last one corresponding to $Z_c$.
We suggest to search for the charged $0^{++}$, $1^{++}$ and $2^{++}$ states in $J/\psi\pi^\pm \pi^0$.
}

\arxivnumber{2404.11215}

\begin{document}

\maketitle

\section{Introduction}

One of the prevailing mysteries in the realm of strong interaction physics is connected to the internal composition of the $XYZ$ states observed by various experiments in the heavy quarkonium mass range (for reviews, see Refs.~\cite{Hosaka:2016pey,Esposito:2016noz,Guo:2017jvc,Olsen:2017bmm,Karliner:2017qhf,Brambilla:2019esw, Yang:2020atz,Chen:2022asf,Meng:2022ozq}). A profound comprehension of this matter could
illuminate the distribution of energy excitation within the nonperturbative domain of quantum chromodynamics (QCD). 
Specifically, it could reveal what kind of clustering (if any) emerges within multiquark states. For example, it could determine whether the quarks cluster
into individual hadrons making the multiquark state a hadronic molecule, or into diquark anti-diquark substructures generating compact states.
The $\X$ state, also known as $\chi_{c1}(3872)$~\cite{ParticleDataGroup:2022pth}, offers an ideal laboratory for addressing these questions.

The $\X$ was observed by the Belle Collaboration~\cite{Belle:2003nnu} in $e^+e^-\to J/\psi\pi^+\pi^-$ and confirmed in many reactions and various final states~\cite{D0:2004zmu, Belle:2006olv, BaBar:2010wfc, BaBar:2010wfc, LHCb:2013kgk, BESIII:2013fnz, ATLAS:2016kwu, BESIII:2019esk, CMS:2021znk, BESIII:2022bse}. Its quantum numbers are $J^{PC}=1^{++}$~\cite{LHCb:2013kgk}. 
The most salient feature of this state is that its mass is very close to the $D^0\bar{D}^{*0}$ threshold---the mass measured from a Flatt\'e analysis by the LHCb Collaboration is $M_X~=~3871.69_{-0.04-0.13}^{+0.00+0.05}$~MeV~\cite{LHCb:2020xds}, while the $D^0\bar D^{*0}$ threshold is at $(3871.69 \pm 0.07)$~MeV~\cite{ParticleDataGroup:2022pth}.
Despite the extremely small phase space, the $X(3872)$ was found to decay into $D^0\bar D^{*0}$ and $D^0\bar D^0\pi^0$ with large branching fractions~\cite{Belle:2008fma,Li:2019kpj,Braaten:2019ags}.
These features suggest that this state is a hadronic molecule~\cite{Close:2003sg,Pakvasa:2003ea,Voloshin:2003nt,Swanson:2003tb,Braaten:2003he,Tornqvist:2004qy} (see Ref.~\cite{Kalashnikova:2018vkv} for a review).
However, there are other interpretations. For instance, the $\X$ was explained as a compact tetraquark state in Ref.~\cite{Maiani:2004vq}. 
Because the quark-gluon interactions are isospin-independent, one distinct prediction in such a tetraquark model is the existence of an isospin triplet as flavor partners of the $\X$. 
The charged partner of the $\X$ was searched for by Belle in the $\jp \pi^+\pi^0$ invariant mass distributions from $B$ meson decays~\cite{Belle:2011vlx}, but no evidence was found.

In this article, based on chiral effective field theory (EFT) with the $\X$ properties as input, we will show in a model-independent way that the $\X$ should have isovector partners (denoted as $W_{c1}$)
also if it is a $D\bar D^*$ hadronic molecule.
They are virtual states\footnote{
A virtual state is defined as a pole below threshold on the real axis on an unphysical Riemann sheet in the strict sense. In this work we will use this term in a broader sense: we  use it also for unphysical sheet poles with a very small imaginary part as long as the path connecting the physical region and the pole has to circle around a threshold.}
of the scattering matrix, and their signals as either a mild cusp or dip at the $D\bar D^*$ threshold should be weak. It should be noted that virtual states can only appear for
hadronic molecular states~\cite{Matuschek:2020gqe}. Thus a confirmation of our claim would at the same time
be a proof for the
molecular nature of the $X(3872)$
as well as its spin partners.

The article is organized as follows. In Section~\ref{sec:sym}, we analyze the consequences of heavy quark spin symmetry and determine the ratio of couplings of the $\X$ to the neutral and charged $D\bar D^*$ channels from experimental measurements.
In Section~\ref{sec:pred}, employing a chiral EFT with $D\bar D \pi$ three-body effects, pole positions of the charged and neutral isovector $W_{c1}$. 
A brief summary is provided in Section~\ref{sec:sum}. 
Technical details of solving the Lippmann-Schwinger equation (LSE) is relegated to Appendix~\ref{app:LSE}, and Appendix~\ref{app:nopion} contains results and discussions in a pionless theory.

\section{Symmetry analysis}
\label{sec:sym}

At energies very close to the $D\bar D^*$ thresholds, we can focus on $S$-wave meson-meson scattering and neglect higher partial waves. Let us first analyze consequences of heavy quark spin symmetry~\cite{Isgur:1989vq}. In the heavy quark limit, heavy quark spins decouple and the total angular momentum of the light degrees of freedom $j_\ell$ becomes a good quantum number. The $D$ and $D^*$ mesons are organized into doublets with $j_\ell^P=1/2^-$, denoted as $H$.
In addition they can be grouped into
isospin doublets,
e.g. $D=(D^+,D^0)^T$.
Thus, the $S$-wave
states containing $H\bar H$ pairs   can
be arranged
to form a definite isospin, $I$, and $j_\ell$.
In particular,  they can be decomposed into four families~\cite{Nieves:2012tt,Guo:2017jvc}, two $(I=0,1)$ with $j_\ell=0$:
\begin{equation}
    0_{\ell I}^{-+} \otimes 0_{c \bar{c}}^{-+}=0_I^{++}, \quad 0_{\ell I}^{-+} \otimes 1_{c \bar{c}}^{--}=1_I^{+-}, \label{eq:jl0}
\end{equation}
and the other two $(I=0,1)$ with $j_\ell=1$:
\begin{equation}
    1_{\ell I}^{--} \otimes 0_{c \bar{c}}^{-+}=1_I^{+-}, \quad 1_{\ell I}^{--} \otimes 1_{c \bar{c}}^{--}=0_I^{++} \oplus 1_I^{++} \oplus 2_I^{++}. \label{eq:jl1}
\end{equation}
Here, the light quark-antiquark pair couples to the charm quark-antiquark pair, $j_{\ell I}^{P C} \otimes s_{c \bar{c}}^{P C}$, to form the $H\bar H$ pairs with quantum numbers $J_I^{P C}$.
Furthermore, $P$ and $C$ denote the parity and charge parity, respectively, and $J$ denotes the total angular momentum of $H\bar H$.
Therefore, for the $\X$ being a $J_I^{PC}=1_0^{++}$ hadronic molecular state, one can predict the existence of three additional isoscalar states with $0_0^{++}$, $1_0^{+-}$, and $2_0^{++}$, respectively, in the strict heavy quark limit~\cite{Hidalgo-Duque:2013pva,Baru:2016iwj}.
In Eq.~\eqref{eq:jl1}, the $J_I^{PC}=0_I^{++}$ and $1_I^{+-}$ states are mixtures of different charmed meson pairs~\cite{Hu:2024hex}, namely,
\begin{equation}
    \begin{aligned}
   J_I^{PC}=0_I^{++}:&\quad \frac{\sqrt{3}}{2} |D \bar{D}\rangle_I +\frac{1}{2}\left|D^* \bar{D}^*\right\rangle_I\, , \\
    J_I^{PC}=1_I^{+-}:&\quad \frac{1}{\sqrt{2}}\left[\left|\left(D \bar{D}^*\right)^{[C=-]}\right\rangle_I+ \left|D^* \bar{D}^*\right\rangle_I \right],
    \end{aligned} \label{eq:mixture}
\end{equation}
respectively, with $\left(D \bar{D}^*\right)^{[C=-]}$ the negative $C$-parity normalized combination of $D\bar D^*$ and $\bar D D^*$.
The $J_I^{PC}=1_I^{++}$ and $2_I^{++}$ states are purely $D\bar D^{*}$ and $D^*\bar D^*$, respectively.
In particular,  a $D^*\bar D^*$ molecule with $2_0^{++}$ is generally expected~\cite{Nieves:2012tt,Hidalgo-Duque:2012rqv,Guo:2013sya,Albaladejo:2015dsa,Baru:2016iwj}, using as input only the $X(3872)$ mass.

One more piece of information that can be used to constrain the $H\bar H$ molecule is the ratio of the branching fractions of the $\X$ decays into $J/\psi\rho^0$ and $J/\psi\omega$, which is from isospin breaking and thus constrains the $I(j_\ell)=1(1)$  sector~\cite{Gamermann:2009fv,Gamermann:2009uq,Hidalgo-Duque:2012rqv,Li:2012cs,Meng:2021kmi,Ji:2022uie}.
The ratio of branching fractions can be used to extract the ratio of the decay amplitudes of the $\X$ decays into $J/\psi\rho^0$ and $J/\psi\omega$~\cite{Suzuki:2005ha,Hanhart:2011tn}, that is, the largely different phase spaces of the two final states are removed from the ratio. The most recent determination by the LHCb Collaboration is~\cite{LHCb:2022jez}
\begin{equation}
    R_X=\left|\frac{\mathcal{M}_{X(3872) \rightarrow J / \psi \rho^0}}{\mathcal{M}_{X(3872) \rightarrow J / \psi \omega}}\right| =0.29 \pm 0.04 \, . \label{eq:RX}
\end{equation}

The $\X$ flavor space wave function as a molecular state is given by
\begin{align}
|X\rangle=\frac{g_{X0}\left|\left(D \bar{D}^*\right)_0^{[C=+]}\right\rangle+g_{X\pm}\left|\left(D \bar{D}^*\right)_\pm^{[C=+]}\right\rangle}{\sqrt{|g_{X\pm}|^2+|g_{X0}|^2}}, \label{eq:X_flavor}
\end{align}
where the first and second term in the numerator are the positive $C$-parity pairs of the neutral 
$(D^0\bar D^{*0}{-}\bar D^0 D^{*0})/\sqrt{2}$ 
and charged $(D^+ D^{*-}{-} D^- D^{*+})/\sqrt{2}$ mesons, respectively,
and $g_{X\pm}$ and $g_{X0}$ are the couplings of the $\X$ to the charged and neutral $D\bar D^*$ pairs.
With these couplings, we have specified the potential deviation from a pure isoscalar wave function.
Being a physical observable, the ratio in Eq.~\eqref{eq:RX} is scale-independent, and is related to  $g_{X\pm}$ and $g_{X0}$ as
\begin{align}
    R_X = \left|\frac{1-R_{\pm/0}}{1+R_{\pm/0}}\right|, \quad R_{\pm/0} \equiv \frac{g_{X\pm}}{g_{X0}}\, .
\end{align}
Here, we have factorized the $X$ decay amplitudes into a long-distance $X\to D\bar D^*$ and a short-distance  $D\bar D^*\to J/\psi V$ part~\cite{Braaten:2005jj}, 
with the long-distance part given by the couplings to the charmed meson pairs. Then the decay amplitudes read,
\begin{align}\nonumber
  \mathcal{M}_{X\to J/\psi V}& =\frac{1}{\sqrt2}\mathcal{M}^{\mathrm{s.d.}(\tilde{\Lambda})}_{[D\bar{D}^{\ast}\to J/\psi V]} \left[g_{X0}L(\tilde{\Lambda},E_0)\pm g_{X\pm}L(\tilde{\Lambda},E_\pm)\right],\\ \nonumber
 & \approx
\frac{1}{\sqrt2}\mathcal{M}^{\mathrm{s.d.}(\tilde{\Lambda})}_{[D\bar{D}^{\ast}\to J/\psi V]}\frac{\mu}{\pi^2}\tilde{\Lambda}     (g_{X0}\pm g_{X\pm})\\ 
&=\frac{1}{\sqrt2}\mathcal{M}^{\mathrm{s.d.}}_{[D\bar{D}^{\ast}\to J/\psi V]}\frac{\mu}{\pi^2}(g_{X0}\pm g_{X\pm}), 
\label{eq:MXJV}
\end{align}
where the loop function is given by 
\begin{equation}
L(\tilde{\Lambda},E_c)=\frac{\mu_c}{\pi^2}\left(\tilde{\Lambda}-\frac{\pi}{2}\sqrt{-2\mu_cE_c}\right)
\end{equation}
with $\mu_c$ the reduced mass in channel $c$,
$E_c$ the difference between the $X$ mass and the threshold of channel $c$,
and $V=\rho^0$ or $\omega$ and the sign is ``$-$'' (``$+$'') for $\rho^0$ ($\omega$), and we have specified in the superscript that the short-distance $\mathcal{M}^{\mathrm{s.d.}(\tilde{\Lambda})}_{[D\bar{D}^{\ast}\to J/\psi V]}$ is cutoff dependent $(\propto 1/\tilde{\Lambda})$, which can absorb the cutoff dependence from the loop at LO~\cite{Braaten:2005jj} resulting in a $\tilde{\Lambda}$ independent short distance amplitude $\mathcal{M}^{\mathrm{s.d.}}_{[D\bar{D}^{\ast}\to J/\psi V]}$. 
To reach the second-to-last line in Eq.~\eqref{eq:MXJV} we used $\mu_0\approx \mu_\pm\approx \mu$ and that the hard momentum scale of the loop is set by the intrinsic scale in the transition $D\bar D^*\to J/\psi V$, which is $\sim \sqrt{2\mu (2M_D-M_{J/\psi})}$, allowing us to neglect not only the tiny binding momentum in the neutral channel $\gamma_0\equiv \sqrt{2\mu_0 \mathcal{B}_X }$ but also
the momentum from the charged channel with $\gamma_\pm=\sqrt{2\mu_\pm E_\pm}\simeq 126$~MeV, where $E_\pm\simeq\Delta$ with
$$\Delta{\equiv} M_{D^\pm}{+}M_{D^{*\pm}} {-} M_{D^0}{-}M_{D^{* 0}}{=}(8.23\pm0.07)~\mbox{MeV} 
$$ employing the masses provided in Ref.~\cite{ParticleDataGroup:2022pth}. 
Since the $\tilde{\Lambda}$ independent short distance $D\bar D^*\to J/\psi V$ amplitudes should be the same up to small flavor SU(3)-breaking corrections for the two decays, 
the short-distance part cancels out
in the ratio $R_X$.
An equivalent relation has been derived in Refs.~\cite{Li:2012cs,Meng:2021kmi}.
Using Eq.~\eqref{eq:RX}, one obtains 
\begin{align}
    R_{\pm/0}=0.55\pm0.05, \label{eq:Rpm0}
\end{align}
which is sizably different from the value obtained
in Refs.~\cite{Gamermann:2009uq,Hidalgo-Duque:2013pva,Albaladejo:2015dsa}, where the ratio in Eq.~\eqref{eq:RX} was assumed to be sensitive to the cutoff-dependent wave function of the $\X$ at the origin.
Since the squared couplings $g_{X\pm}^2$ and $g_{X0}^2$ correspond to the residues of the $\X$ pole in the elastic scattering amplitudes of the charged and neutral meson pairs, respectively, the ratio $R_{\pm/0}$ and the $\X$ mass as two inputs can be employed to constrain the $j_\ell=1$ $S$-wave $H\bar H$ interactions in both the isoscalar and isovector channels.

\section{Predictions}
\label{sec:pred}

Similar to the matured chiral EFT for nucleon-nucleon interaction~\cite{Epelbaum:2008ga}, one can formulate a chiral EFT for the $D\bar D^*$ systems. The $D\bar D^*$ interactions can be described as a sum of contact terms and one-pion exchange (OPE) derived from the $D^*D\pi$ chiral Lagrangian~\cite{Wise:1992hn}. 

Because the $\X$ is close to the $D^0\bar D^{*0}$ threshold, the contact terms can be expanded in powers of a small parameter $Q/\Lambda$, where $\Lambda$ is a hard scale, and $Q$ is the momentum scale of the charmed meson, which may be identified with $\gamma_\pm$ provided in the 
previous section.

At leading order (LO), there are two low-energy constants (LECs), $C_{0X}$ and $C_{1X}$, in the isoscalar and isovector channels, respectively~\cite{Hidalgo-Duque:2012rqv}.
\begin{align}  V_{\mathrm{ct}}=\frac{1}{2}
\begin{pmatrix}[1.5] C_{0X}{+}C_{1X} & C_{0X}{-}C_{1X} \\ C_{0X} {-}C_{1X} & C_{0X} {+}C_{1X} \\
\end{pmatrix},
\end{align}
where the matrix is in the channel space with the first and second channels being the neutral and charged $D\bar D^*$ pairs, respectively, specified in Eq.~\eqref{eq:X_flavor}.

\begin{figure}[tb]
    \centering
    \includegraphics[width=0.68\linewidth]{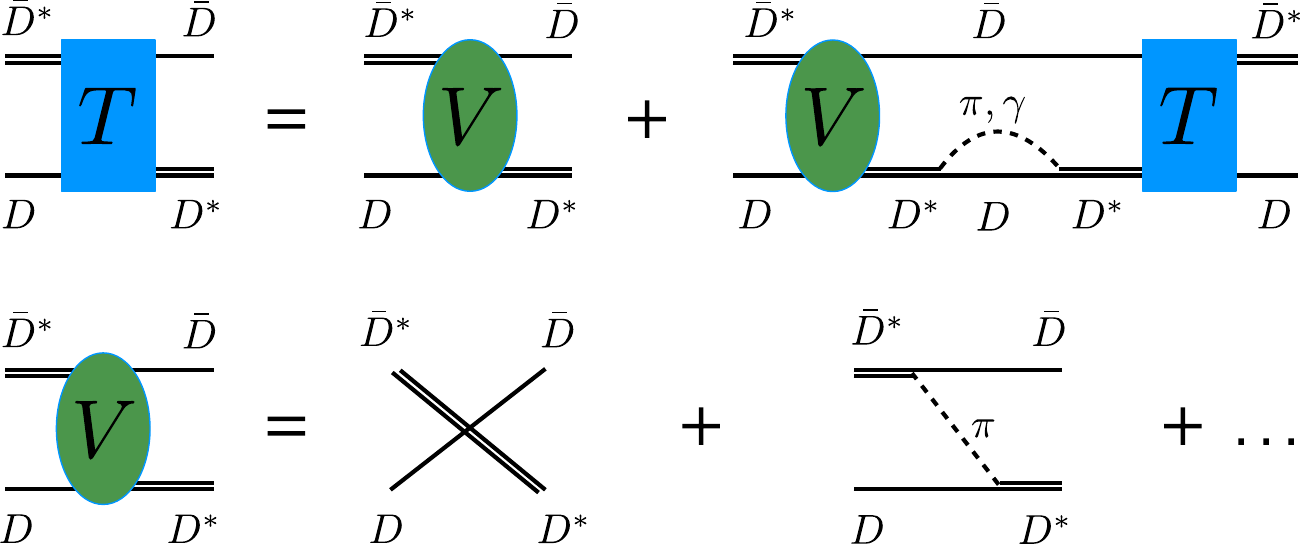}
    \caption{$D\bar D^*$ scattering amplitudes with the full $D\bar D\pi$ three-body effects. }
    \label{fig:Tmatrix}
\end{figure}
The amplitudes with the full $D\bar D\pi$ three-body effects are solutions of the LSE, as shown in Fig.~\ref{fig:Tmatrix},
\begin{align}
    T(E;p^\prime,p)= V(E;p^\prime,p) +\int\frac{l^2dl}{2\pi^2}V(E;p',l)G(E;{l})T(E;l,p), \label{eq:LSE}
\end{align}
with $E\equiv\sqrt{s}-M_{D^0}-M_{D^{\ast 0}}$, $\sqrt{s}$ the total energy in the center-of-mass (c.m.) frame, $p^{(\prime)}$ the magnitude of the initial (final) state c.m. three-momentum, $G$ the Green's function for the
two heavy mesons, and $V$ the scattering potential including both the contact term and the OPE,
\begin{align}
V(E;p^\prime,p)=V_{\mathrm{ct}}+V_{\pi}(E;p^\prime,p)\, .
\end{align}
The $D^*$ widths are considered in $G$
with both the $D\pi$ and $D\gamma$ contributions included. 
We treat the pions dynamically, approximate the photon loops in Fig.~\ref{fig:Tmatrix} by imaginary constants determined from the $D^*$ radiative decay widths, and neglect the photon exchange potential, which binds the $X$ atom as a $D^\pm D^{*\mp}$ Coulombic bound state~\cite{Zhang:2020mpi} but does not affect the formation of hadronic molecular systems.
Explicit expressions for the Green's function, the OPE potential, and technicalities regarding the choice of 
the integration
path~\cite{Dong:2024fjk} can be found in Appendix~\ref{app:LSE}. 

\begin{figure}[tb]
    \centering
    \includegraphics[width=0.75\linewidth]{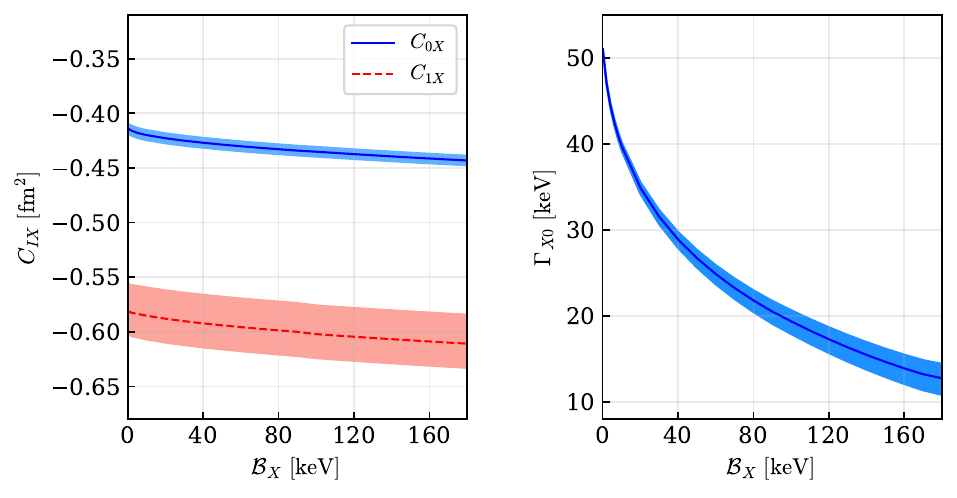}
    \caption{Left: $C_{0X}$ and $C_{1X}$ at $\Lambda=0.6$~GeV as functions of $\mathcal{B}_X$. Right: open-charm partial width of $\X$ from the $D\bar D\pi$ and $D\bar D\gamma$ modes through $D^*$ and $\bar D^*$ decays. The bands are due to the uncertainty in Eq.~\eqref{eq:Rpm0}.}
    \label{fig:LECs}
\end{figure}

The LSE is regularized with a momentum cutoff $\Lambda$ and
solved numerically~\cite{Doring:2009yv,Baru:2011rs, Du:2021zzh,Ji:2022blw,Dong:2024fjk}.
By requiring the real part of the $\X$ pole to be $E=-\mathcal{B}_X$,
where the binding energy is defined as $\mathcal{B}_X \equiv M_{D^0}+M_{D^{* 0}}-M_{X(3872)}$,  and 
\begin{align}
    R_{\pm/0} = \lim_{E\to -\mathcal{B}_X-i\Gamma_{X0}/2} \frac{T_{21}(E)}{T_{11}(E)},
\end{align}
with $\Gamma_{X0}$ the open-charm partial width of $\X$ from the $D\bar D\pi$ and $D\bar D\gamma$ modes through the $D^*$ and $\bar D^*$ decays, we can determine $C_{0X}$ and $C_{1X}$.
The solution was found employing an iterative procedure: We use the $D^{*0}$ width 55.3~keV~\cite{Guo:2019qcn} as the initial value for $\Gamma_{X0}$,
then $C_{0X}$ and $C_{1X}$ 
are fixed to reproduce both the real part of the $\X$ pole and the coupling ratio in Eq.~\eqref{eq:Rpm0}. From those LECs $\Gamma_{X0}$ is recalculated, and the procedure is repeated
until convergence is achieved. The resulting $\Gamma_{X0}$ is shown in the right panel of Fig.~\ref{fig:LECs},
and the LECs fixed in this way in the left panel for  $\Lambda=0.6$~GeV.

Similar to the case of the pionfull EFT treatment of the $T_{cc}(3875)$ in Ref.~\cite{Du:2021zzh}, the smallness of the expansion parameter $Q/\Lambda$ ensures that the momentum independent contact terms can absorb the LO cutoff dependence, with the remainder a higher order effect relatively suppressed by $\mathcal{O}(Q^2/\Lambda^2)$.
We have checked that varying the cutoff from 0.5 to 1.0~GeV, the predicted pole positions for the spin partner states (relative to the corresponding thresholds) change less than 5\%.

Since all the $S$-wave $H\bar H$ meson interactions for pairs listed in Eq.~\eqref{eq:jl1} depend on the same $C_{1X}$, we can predict in a model-independent way the existence of isovector $H\bar H$ hadronic molecules using the $C_{1X}$ value given in Fig.~\ref{fig:LECs}.\footnote{The predictions in Ref.~\cite{Ji:2022uie} are cutoff dependent as the cutoff dependence in the $X\to J/\psi\rho$ and $J/\psi\omega$ amplitudes was not properly renormalized.}
In particular, we can robustly predict the $J_I^{PC}=1_1^{++}$ ($W_{c1}$) states as isovector $(D\bar D^*)^{[C=+]}$ hadronic molecules.
With the notation in Eq.~\eqref{eq:X_flavor}, the flavor wave function of the neutral $W_{c1}$ is
\begin{align}
    \left|W_{c 1}^0\right\rangle=\frac{g_{W0}\left|\left(D \bar{D}^*\right)_0^{[C=+]}\right\rangle+g_{W\pm}\left|\left(D \bar{D}^*\right)_\pm^{[C=+]}\right\rangle}{\sqrt{|g_{W\pm}|^2+|g_{W0}|^2}},
    \label{eq:Wc1_flavor}
\end{align}
where $g_{W\pm}$ and $g_{W0}$ are the couplings of the $W_{c1}^0$ to the charged and neutral $D\bar D^*$ pairs, respectively.
Despite that the absolute value of $C_{1X}$ is larger than that of $C_{0X}$ at $\Lambda= 0.6$~GeV, the inclusion of the OPE leads to a weaker attraction in the isovector channel than the isoscalar one.
Consequently, 
while the $\X$ is a bound state pole 
in the stable $D^*$ limit,
the charged $W_{c1}^\pm$ is  a virtual state pole in the stable $D^*$ limit, but of a single-channel $T$ matrix.
The pole location of the neutral $W_{c1}^0$, being in the same coupled-channel $T$ matrix in Eq.~\eqref{eq:LSE} as the $\X$, is more complicated. It would be a virtual state pole like the charged $W_{c1}^\pm$ were there a single channel; however, the $(D\bar D^{*})_0^{[C=+]}${-}$(D\bar D^{*})_\pm^{[C=+]}$ channel coupling applies an effective repulsion to the higher channel, pushing the pole from below to above the $D^+D^{*-}$ threshold on the fourth Riemann sheet (RS), i.e., RS$_{+-}$~\cite{Zhang:2024qkg} (for a definition of RSs, see Appendix~\ref{app:RSs}). 
\begin{figure}[tb]
    \centering
    \includegraphics[width=0.6\linewidth]{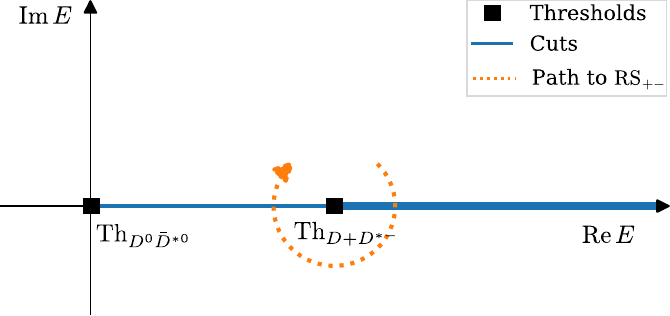}
    \caption{Path from the physical region to RS$_{+-}$ in the complex $E$ plane. Th$_{D^0\bar D^{*0}}$ and Th$_{D^+\bar D^{*-}}$ denote the ${D^0\bar D^{*0}}$ and ${D^0\bar D^{*0}}$ thresholds, respectively.}
    \label{fig:rs}
\end{figure}
The pole on RS$_{+-}$ can only be reached from the physical region by circling around the $D^+D^{*-}$ threshold as shown in Fig.~\ref{fig:rs}.
Consequently, the pole should manifest as a threshold cusp, just like a usual virtual state. 

\begin{figure}[tb]
    \centering
    \includegraphics[width=0.618\linewidth]{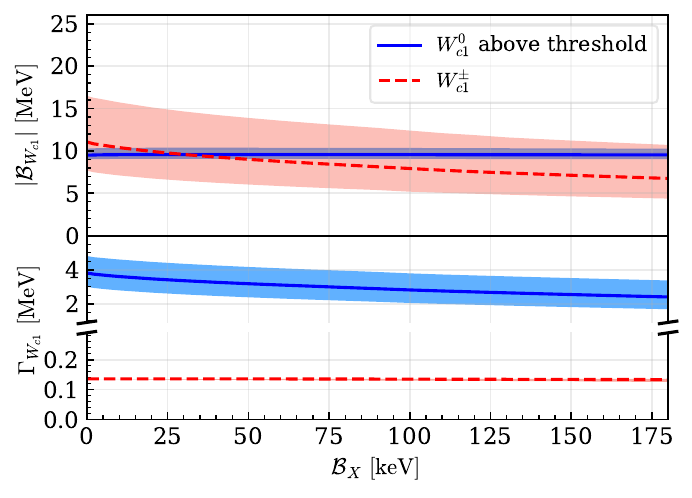}
    \caption{Dependence of the predicted $W_{c1}$ pole positions on the input $\X$ binding energy. The bands are from the uncertainty in Eq.~\eqref{eq:Rpm0}. Upper: magnitude of the real part of the pole position with respect to the $D^0\bar D^{*0}$ (for $W_{c1}^0$, above threshold) or $D^0 D^{*-}$ (for $W_{c1}^-$, below threshold) threshold.
    Lower: twice the magnitude of the imaginary part of the $W_{c1}$ pole position. }
    \label{fig:Wc1_poles}
\end{figure}
In Fig.~\ref{fig:Wc1_poles}, we show the pole positions of the predicted $W_{c1}^0$ and $W_{c1}^\pm$ virtual states, which are\footnote{These poles would become virtual state poles below threshold on the real axis of the unphysical RS in the single-channel case without three-body effects.}
\begin{align}
  \begin{aligned}
    {W_{c1}^0:} &\quad  {3881.2^{+0.8}_{-0.0}- i 1.6^{+0.7}_{-0.9}~{\rm MeV},} \\    
    W_{c1}^\pm: &\quad  3866.9^{+4.6}_{-7.7}- i (0.07\pm0.01)~{\rm MeV},
  \end{aligned}
\end{align}
{The $W_{c1}^0$ pole is about $9.5^{+0.8}_{-0.0}$~MeV above the $D^0\bar D^{*0}$ threshold and $1.3^{+0.8}_{-0.0}$~MeV above the $D^+D^{*-}$ threshold, while the $W_{c1}^{-}$ is located $8_{-5}^{+8}$~MeV below the $D^0D^{*-}$ threshold.
We also find another pole at $3865.3^{+4.2}_{-7.4}- i 0.15^{+0.04}_{-0.03}~{\rm MeV}$ on the third RS (RS$_{--}$) in the neutral coupled-channel $T$ matrix. Although its isovector residue is much larger than the isoscalar one, it is a shadow pole~\cite{Eden:1963zz} of the $X(3872)$ since they coincide if we switch off the channel coupling.
This shadow pole hardly has any physical significance as it is shielded by the $X(3872)$ and the two $D^0\bar D^{*0}$ and $D^+D^{*-}$ thresholds.
For a detailed discussion about all the poles in a two-channel system and their evolution along with changing the interaction and coupling strengths, we refer to Ref.~\cite{Zhang:2024qkg}.
}

\begin{figure}[tb]
    \centering
    \includegraphics[width=0.618\linewidth]{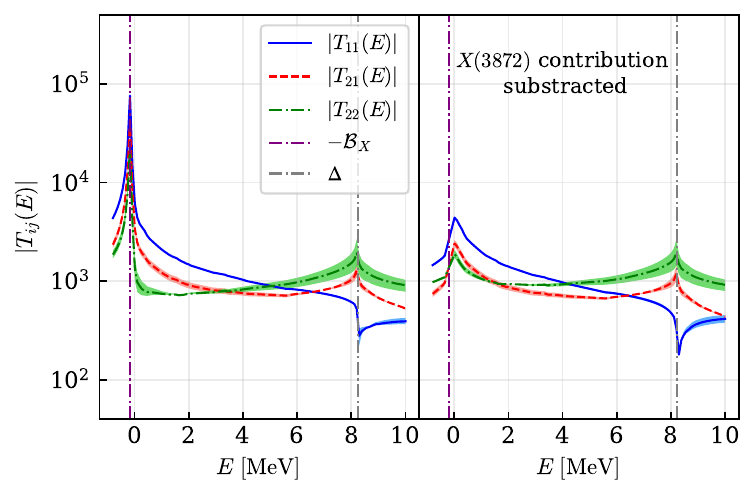}\hfill
    \includegraphics[width=0.618\linewidth]{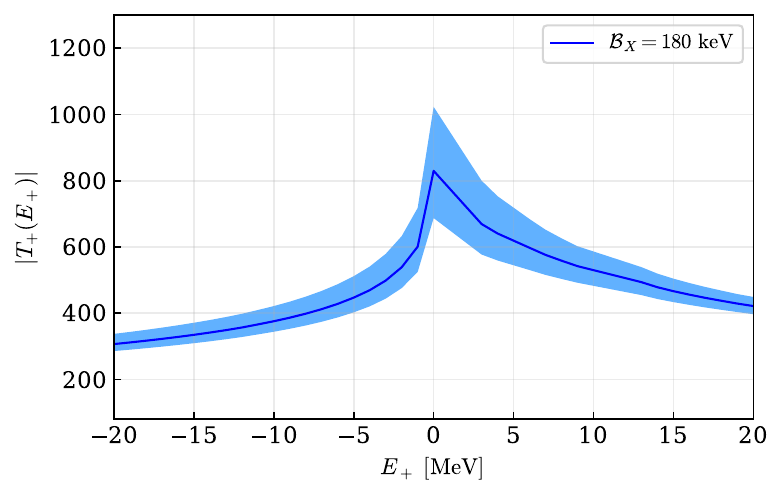}
    \caption{Upper: Line shapes of the $S$-wave $D^0\bar D^{*0}$-$D^+D^{*-}$ scattering $T$-matrix elements, where the left and right parts are for the full $T$-matrix elements and the $T$-matrix elements with the $X(3872)$ pole contribution subtracted, respectively. 
    Lower: Line shape of the single-channel $D^+\bar D^{*0}$ scattering $T$-matrix element, with $E_+$ defined relative to the $D^+\bar D^{*0}$ threshold.
    Here $\mathcal{B}_X$ is set to 180~keV and the uncertainty is from Eq.~\eqref{eq:Rpm0}.
    }
    \label{fig:lineshape_full}
\end{figure}

The {$W_{c1}^{\pm}$} can decay into $J/\psi\pi\pi$ without breaking isospin symmetry, and the neutral one {could} decay into $J/\psi\pi^+\pi^-$, which is the discovery channel of the $\X$~\cite{Belle:2003nnu}. Thus, one natural question is why its signal and the signal of the charged $W_{c1}^\pm$ states have not been observed~\cite{Belle:2011vlx}.
To answer this question, we show the line shapes
that emerge from the $(D\bar D^{*})_0^{[C=+]}${-}$(D\bar D^{*})_\pm^{[C=+]}$ coupled-channel
$T$ matrix and the isovector $D^+\bar D^{*0}$ single-channel scattering amplitude in Fig.~\ref{fig:lineshape_full}, where $\mathcal{B}_X=180$~keV is used as input for the $X(3872)$ binding energy.
Being virtual states, the $W_{c1}$ appear as threshold cusps. In the upper plot, the $W_{c1}^0$ signal is partly hidden under the $\X$ signal
which is represented
by the huge narrow peak around the $D^0\bar D^{*0}$ threshold.
The height of this peak depends crucially on the $\X$ binding energy. For $\mathcal{B}_X>0$, the maximum is located at $-\mathcal{B}_X$. In that case, since the peak is below threshold, the maximal $|T_{ij}|$ values are larger than the values at the $D^0\bar D^{*0}$ threshold. It is also more than one order of magnitude larger than the cusps at the $D^+D^{*-}$ threshold as well as the cusp in the $D^+\bar D^{*0}$ channel, explaining why so far no significant signal around 3880~MeV has been reported.
Furthermore, subtracting out the $\X$ contribution significantly reduces the peak at the lower threshold and modifies the structure at the higher. 
The line shape of $|T_{ij} - g_{Xi}g_{Xj}/(E + \mathcal{B}_X + i\Gamma_{X0}/2)|$, with $g_{Xi}$ the coupling of $X(3872)$ to channel-$i$, is shown in the right panel of the upper plot in Fig.~\ref{fig:lineshape_full}. 
It is evident that  the remaining line shapes possess threshold cusps of similar strengths at both the $D^0\bar D^{*0}$ and $D^+D^{*-}$ thresholds.

The dip in the $|T_{11}(E)|$ line shape at the threshold of charged charmed meson pair is due to the strong $S$-wave attraction of the $(D\bar D^{*})_\pm^{[C=+]}$ pair. It is another manifestation of the mechanism introduced in Ref.~\cite{Dong:2020hxe}, which also offers a natural explanation~\cite{Baru:2024ptl} of the dip around the $\X$ mass in $e^+e^-$ direction production cross section measured by BESIII~\cite{BESIII:2022ner}.

However, if the $\X$ approaches the zero binding limit, that is $\mathcal{B}_X\to 0$, the residues of the $T$-matrix elements at the $\X$ pole will be significantly reduced since they scale as $\sqrt{2\mu_0\mathcal{B}_X}$.\footnote{No such relation exists for virtual states.}
Then removing the $\X$ pole contribution would hardly change the line shapes, and the $W_{c1}$ signals would be relatively more prominent, in conflict with experiment.
Therefore, we conclude from the above analysis that the $\X$ must be
either a bound state with a non-vanishing binding energy or a very close to threshold virtual state
in order to prevent a strong signal from the isovector $W_{c1}$ states.

In the strict heavy quark limit, when $D$ and $D^*$ are degenerate, the $W_{c1}$ should have isovector heavy quark spin partners with $J_I^{PC}=0^{++}_1$, $2^{++}_1$ and $1_1^{+-}$ listed in Eq.~\eqref{eq:jl1}.
As shown in  Eq.~\eqref{eq:mixture}, the first one involves both $S$-wave $D\bar D$ and $D^*\bar D^*$, and the third one, which, mixing with the $1_1^{+-}$
state
from  Eq.~\eqref{eq:jl0},
form
 the $Z_c(3900)$~\cite{BESIII:2013ris,Belle:2013yex}
 and the $Z_c(4020)$~\cite{BESIII:2013ouc,BESIII:2013mhi}, involves both $S$-wave $D\bar D^*$ and $D^*\bar D^*$. In addition, the second one can decay into $D\bar D$ and $D\bar D^*$ in $D$-waves.
Hence, a full calculation of all these cases would involve coupled channels with threshold differences of about 0.14 or 0.28~GeV, which relate to
quite large momentum scales of the
order of 0.5 and 0.7~GeV, respectively, requiring the
inclusion of higher terms in the chiral expansion to reach sufficient accuracy. 
Moreover, a proper treatment of 
the then unavoidable $D$-wave contributions needs
an $S$--$D$ counterterm promoted to LO to render the results regulator independent~\cite{Wang:2018jlv,Baru:2019xnh}. To fix this counterterm, additional data need to be included.
Thus, we discuss these sectors more qualitatively in terms of a pionless theory.
The LECs $C_{0X}$ and $C_{1X}$ are determined similarly as above. We checked that in the pionless theory the features of the
results discussed above
remain (for details, see Appendix~\ref{app:nopion}). For example,
a virtual state pole is obtained for $W_{c1}$  by about 10~MeV below threshold. Varying the two-meson reduced mass between $M_D/2$ and $M_{D^*}/2$, the virtual energy changes only by $\pm3$~MeV. 
Therefore, we anticipate virtual states in all the  $0^{++}_1$ ($W_{c0}$), $1^{+-}_1$ ($Z_{c}$), and $2^{++}_1$ ($W_{c2}$) channels.
Yet, since a coupling to higher channels effectively provides an additional attraction, the first two virtual states could be closer to threshold or even become bound state(s).
Similarly, one expects the $S$-wave $D_s \bar{D}^*-D_s^* \bar{D}$ with $J^P=1^+$ and  $D_s^* \bar{D}^*$ with $J^P=2^+$, which share the same  LEC $C_{1X}$ as the isovector $W_{c1}$, to have virtual state poles and thus show up as threshold cusps (for a list of LECs for all $S$-wave $D_{(s)}^{(*)}\bar D_{(s)}^{(*)}$ pairs and predictions using a constant-contact-term potential, we refer to Tables III, IV and V in Ref.~\cite{Ji:2022uie}).

Experimentally hidden charm tetraquarks with strangeness, traditionally called $Z_{cs}$, are reported in Refs.~\cite{BESIII:2020qkh,LHCb:2021uow}, but no sound conclusion about the pole locations of the
states can be drawn from either of them. In particular, while a candidate for the above mentioned state is reported in both experiments close to the $D_s\bar D^*$ and the $\bar D_s^*D$ thresholds with comparable mass,  strikingly different widths were extracted 
in the experimental analyses. The former data were analysed in Refs.~\cite{Yang:2020nrt,Baru:2021ddn, Ortega:2021enc, Du:2022jjv}, however, given the current quality of data a definite conclusion regarding the location of the pertinent pole is not yet possible, especially since the structure in the data emerges as a subtle interplay of the pole and triangle singularity~\cite{Yang:2020nrt,Baru:2021ddn, Du:2022jjv}.
In Refs.~\cite{Nakamura:2021bvs,Luo:2022xjx} on the other hand the LHCb data~\cite{LHCb:2021uow}
was shown to be consistent with the presence of only kinematic singularities without the
need to include any poles.

In Ref.~\cite{Maiani:2021tri} it is claimed that the emergence of SU(3) multiplets are a unique feature of the compact tetraquark models. However, 
since SU(3) flavor is an approximate symmetry of QCD it 
naturally leads an imprint also in molecular structures as discussed in this work as well as earlier, see, e.g., Ref.~\cite{Ji:2022uie}.

\section{Summary}
\label{sec:sum}

In this article, we show that the $\X$ treated
as an isoscalar
$D\bar D^*$ hadronic molecule,
must have an isovector partner
with the same particle
content and $J_I^{PC}=1_1^{++}$, based on chiral EFT 
employing only $\X$ properties as input. The predicted states are virtual states in the scattering matrix that leave
an imprint in the data either as a mild cusp or dip at the $D\bar D^*$ threshold. 
The pole positions are at $3865.3^{+4.2}_{-7.4}- i 0.15^{+0.04}_{-0.03}$~MeV for $W_{c1}^0$ and  $3866.9^{+4.6}_{-7.7}- i (0.07\pm0.01)$~MeV for $W_{c1}^\pm$.
Moreover,
the states should have isovector heavy quark spin partners with $J_I^{PC}=2_1^{++}$ ($W_{c2}$), $1_1^{+-}$ (corresponding to a mixture of $Z_c(3900)$ and $Z_c(4020)$),
and $0_1^{++}$ ($W_{c0}$).
All these are expected to be virtual states, though channel couplings to $D^*\bar D^*$ in some cases may promote a virtual state to a bound one.
While isovector partner states are expected also for compact tetraquarks~\cite{Maiani:2004vq}, 
only a molecular nature allows for
virtual states~\cite{Matuschek:2020gqe}. Virtual states leave an imprint in the data exactly at threshold which is thus a model-independent prediction from the molecular scenario discussed in this article.

The findings here imply that the resonant peak in the $J/\psi\pi^+\pi^-$ distributions around 3872~MeV is not purely from the $\X$ but also contains contributions from the $W_{c1}^0$ state.
The charged $W_{c1}^\pm$ state should lead to a cusp at the $(D\bar{D}^{*})^\pm$ threshold, and can be searched for in high-statistics data samples of the $J/\psi\pi^\pm \pi^0$.
The charged $W_{c0}^\pm$ and $W_{c2}^\pm$, whose mass predictions are more uncertain, may also be searched for in the same final state.

\bigskip

\begin{acknowledgments}
We are grateful to Alexey Nefediev and Bing-Song Zou for useful discussions. ZHZ would like to thank Zhao-Sai Jia for checking the numerical results.
This work is supported in part by the National Key R\&D Program of China under Grant No. 2023YFA1606703; by the Chinese Academy of Sciences (CAS) under Grants No.~YSBR-101 and No.~XDB34030000;
by the National Natural Science Foundation of China (NSFC) under Grants No. 12125507, No. 12361141819, and No. 12047503; and by NSFC and the Deutsche Forschungsgemeinschaft (DFG) through the funds provided to the Sino-German Collaborative Research Center CRC110 ``Symmetries and the Emergence of Structure in QCD'' (DFG Project-ID 196253076). A.R. and U.-G.M., in addition, thank the CAS President's International Fellowship Initiative (PIFI) (Grants No. 2024VMB0001 and No. 2025PD0022, respectively) for partial financial support. 
\end{acknowledgments}

\begin{appendix}

\section{Details of the Lippmann-Schwinger equation}
\label{app:LSE}

The introduction of a dynamical pion in the interaction between $D\bar D^*$ leads to the emergence of nontrivial cuts in the $D\bar D^*$ scattering amplitude in two ways~\cite{Baru:2011rs, Du:2021zzh,Ji:2022blw,Dong:2024fjk}: first, the on-shellness of the exchanged pion between $D\bar D^*$, depicted in Fig.~\ref{fig:3body_cut} (a); and second, the energy-dependent width of $D^*$, depicted in Fig.~\ref{fig:3body_cut} (b). These two aspects are discussed in detail here.

The $D^*D\pi$ coupling can be described by the Lagrangian~\cite{Wise:1992hn},
\begin{align}
\mathcal L_{DD^*\pi}=-\frac{\sqrt 2g}{F_\pi}\sqrt{m_Dm_{D^*}}\left\{\left[D(\partial^\mu \mathcal P) D^{*\dagger}_\mu+D_\mu^*(\partial^\mu \mathcal P) D^{\dagger}\right]-\left[\bar D(\partial^\mu \mathcal P)^T \bar D^{*\dagger}_\mu+\bar D_\mu^*(\partial^\mu \mathcal P)^T \bar D^{\dagger}\right]\right\},
\label{eq:LDsDpi}
\end{align}
where $F_{\pi}=92.1~\text{MeV}$ is the pion decay constant, $g=0.57$ is determined by the $D^*\to D\pi$ decay width~\cite{ParticleDataGroup:2022pth}, and 
\begin{align}
    \begin{aligned}
&D=(D^+,D^0),\qquad\bar D=(D^-,\bar D^0),\\
&D^*=(D^{*+},D^{*0}),~\bar D^*=(D^{*-},\bar D^{*0}),\\
&\mathcal P=\left(\begin{array}{cc}
\pi^0/\sqrt2&\pi^+\\
\pi^-&-\pi^0/\sqrt2
\end{array}
\right).
\end{aligned}
\end{align}
We have used the following phase conventions for charge conjugation,
\begin{align}
    \hat C|D\rangle=|\bar D\rangle,\quad\hat C|D^*\rangle=-|\bar D^*\rangle.
\end{align}
    
\subsection{OPE}

\begin{figure}[tb]
        \centering
        \includegraphics[width=0.68\linewidth]{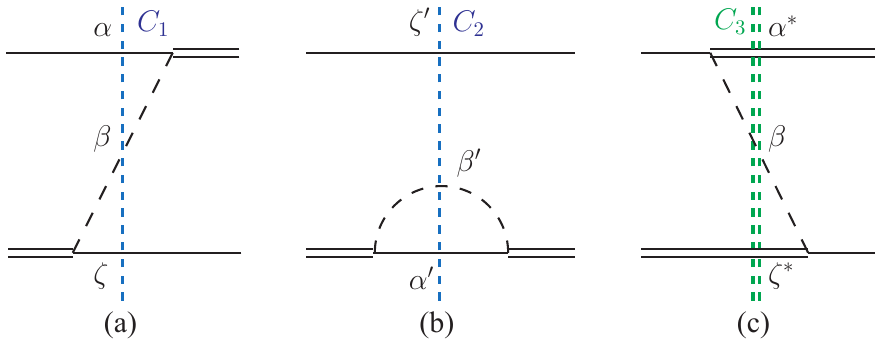}
        \caption{Illustration of the 3-body cuts in the (a) retarded propagator, (b) $D^\ast$ selfenergy, and (c) advanced propagator. The $D^\ast$, $D$,
        and $\pi$ mesons are represented by the double solid, solid, and dashed lines, respectively. The three particles on the blue cut $C_1$ or $C_2$ can go on-shell in the physical energy region, while the three particles on the green cut $C_3$ cannot. The labels of the particles on the cuts are those in Eqs.~\eqref{eq:propagator_TOPT} and \eqref{eq:pion_momentum_cubic_in_width}.}
        \label{fig:3body_cut}
\end{figure}

Using the Lagrangian in Eq.~\eqref{eq:LDsDpi}, the OPE potential of the $D^0\bar D^{*0}$-$D^+D^{*-}$ coupled channels in $S$-wave can be expressed as
    \begin{align}
        V_{\pi}(E;p^\prime,p)=\frac{g^2}{6F_{\pi}^2}
        \begin{pmatrix}[1.5]
            \frac{1}{2}V^{SS}_{D^0\pi^0\bar{D}^{0}} & V^{SS}_{D^0\pi^+D^-} \\
            V^{SS}_{D^+\pi^-\bar{D}^0} &\frac{1}{2}V^{SS}_{D^+\pi^0D^-}
        \end{pmatrix},
    \end{align}
    with
     \begin{align}
        V^{SS}_{\alpha\beta\zeta}=\frac{1}{2}\int_{-1}^{1}dz \frac{q^2}{D_{\alpha\beta\zeta}^{\mathrm{TOPT}}(E;p^\prime,p,z)},
        \label{eq:S_wave_projection}
    \end{align}
    and $q^2=p^{\prime 2}+p^2-2p^\prime p z$.
The propagator of the exchanged pion reads  in time-ordered perturbation theory (TOPT) as
    \begin{align}
    \frac{1}{D_{\alpha\beta\zeta}^{\mathrm{TOPT}}}=\frac{1}{2E_{\beta}(q)}\left(\frac{1}{D_{\alpha\beta\zeta}^R}+\frac{1}{D_{\alpha^*\beta \zeta^*}^A}\right),
        \label{eq:propagator_TOPT}
    \end{align}
    with
    \begin{align}
        D_{\alpha\beta\zeta}^R&=\sqrt{s} - E_{\alpha}(p) -E_{\beta}(q)-E_{\zeta}(p')+i\varepsilon,\\
        D_{\alpha^*\beta \zeta^*}^A&=\sqrt{s} - E_{\alpha^*}(p') -E_{\beta}(q)-E_{\zeta^*}(p)+i\varepsilon.
    \end{align}
    These two terms correspond to the diagrams in Fig.~\ref{fig:3body_cut} (a) and (c), respectively. Here the energy for particle $\alpha$ with 3-momentum magnitude $l$ is expressed as
    $E_{\alpha}(l)=\sqrt{l^2+M_{\alpha}^2}$.
    The labels $\alpha^\ast$ and $\zeta^\ast$ represent $D^\ast$ and $\bar{D}^\ast$, respectively, and their electric charges are given by 
    \begin{align}
        Q_e(\alpha^\ast)=Q_e(\alpha)+Q_e(\beta), \quad Q_e(\zeta^\ast)=Q_e(\beta)+Q_e(\zeta).
    \end{align} 

    \begin{figure}[tb]
    \centering
    \includegraphics[width=0.75\linewidth]{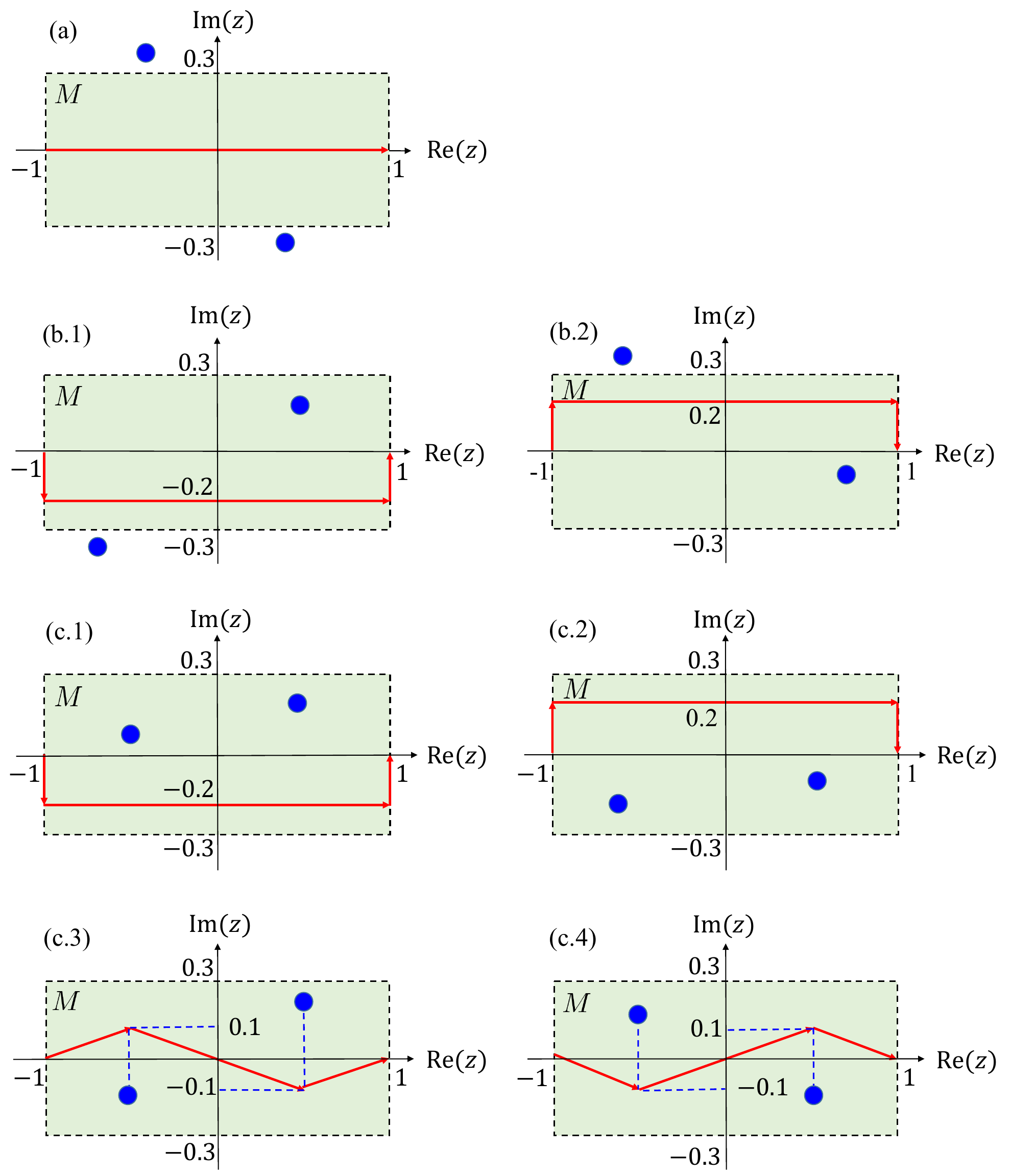}
    \caption{Possible positions of the singularities $z_{0R}$ and $z_{0\pi}$ (blue dots) relative to region $M$ (shaded area) and the corresponding integration paths (in red).}
    \label{fig:integrate_path}
\end{figure}

{The integrand of Eq.~\eqref{eq:S_wave_projection} is divergent at the roots $z_{0R}$ and $z_{0\pi}$ of $D^R_{\alpha \beta \zeta}(E;p^\prime,p,z)$ and $E_{\beta}(q)$, respectively. One should deform the integration
path to avoid the singularities being too close to it.
We use the strategy
in Ref.~\cite{Dong:2024fjk}. That is, we define a neighbourhood $M=\Big\{z\in \mathbb{C}\Big|-1<\mathrm{Re}[z]<1,-0.3<\mathrm{Im}[z]<0.3\Big\}$ of $z\in[-1,1]$ and choose the integration path properly according to the relative positions of $z_{0R}$ and $z_{0\pi}$ to $M$. Without loosing generality, we assume $\mathrm{Re}[z_R]\geq \mathrm{Re}[z_\pi]$ in the following discussion on the choice of the integration path. There are several cases:
\begin{enumerate}
   \item[(a)] $z_{0R}\notin M\; \& \;z_{0\pi}\notin M$. Both singularities are not in $M$. The integration path is just 
   \begin{align}
        C_a=\Big\{z\in \mathbb{R}\Big|-1\leq z\leq 1\Big\},
   \end{align}
   like (a) in Fig.~\ref{fig:integrate_path}.
   \item[(b)] $\big( z_{0R}\in M\; \& \; z_{0\pi}  \notin M \big)\;\big|\big|\;\big( z_{0R} \notin M\; \& \; z_{0\pi} \in M\big) $. Only one of the singularities is located in $M$ and is denoted as $z_M$. The path depends on the position of $z_M$ in $M$.
    \begin{enumerate}
        \item[(1)] $\mathrm{Im}[z_M]>0$. The path is chosen as
        \begin{align}
            C_{b1}&=C_{b1}^{(1)}\cup C_{b1}^{(2)} \cup C_{b1}^{(3)},\\
            C_{b1}^{(1)}&=\Big\{z(t)\in \mathbb{C}\Big|z(t)=-1-it/5, 0 \leq t \leq 1\Big\}, \\
            C_{b1}^{(2)}&=\Big\{z(t)\in \mathbb{C}\Big|z(t)=-1-i/5+2t, 0 \leq t \leq 1\Big\}, \\
            C_{b1}^{(3)}&=\Big\{z(t)\in \mathbb{C}\Big|z(t)=1-i(1-t)/5, 0 \leq t \leq 1\Big\},
        \end{align}
        like (b.1) in Fig.~\ref{fig:integrate_path}.
        \item[(2)] $\mathrm{Im}[z_M]<0$.
        The path is chosen as
        \begin{align}
            C_{b2}&=C_{b2}^{(1)}\cup C_{b2}^{(2)}\cup C_{b2}^{(3)},\\
            C_{b2}^{(1)}&=\Big\{z(t)\in \mathbb{C}\Big|z(t)=-1+it/5, 0 \leq t \leq 1\Big\}, \\
            C_{b2}^{(2)}&=\Big\{z(t)\in \mathbb{C}\Big|z(t)=-1+i/5+2t, 0 \leq t \leq 1\Big\}, \\
            C_{b2}^{(3)}&=\Big\{z(t)\in \mathbb{C}\Big|z(t)=1+i(1-t)/5, 0 \leq t \leq 1\Big\},
        \end{align}
        like (b.2) in Fig.~\ref{fig:integrate_path}.
    \end{enumerate}
    \item[(c)] $ z_{0R}\in M\; \& \; z_{0\pi}  \in M $. Both singularities are in the region $M$.
    The are four cases of the singularity positions.
    \begin{enumerate}
        \item[(1)] $\mathrm{Im}[z_R]>0\;\& \;\mathrm{Im}[z_\pi]>0$. Both singularities are on the upper half $z$ plane. The path is chosen as $C_{c1}=C_{b1}$, like (c.1) in Fig.~\ref{fig:integrate_path}.
        \item[(2)] $\mathrm{Im}[z_R]<0\;\& \;\mathrm{Im}[z_\pi]<0$. Both singularities are on the lower half $z$ plane. The path is chosen as
        $C_{c2}=C_{b2}$, like (c.2) in Fig.~\ref{fig:integrate_path}.
        \item[(3)] $\mathrm{Im}[z_R]>0\;\& \;\mathrm{Im}[z_\pi]<0$. The two singularities are on the opposite side of the real axis on the $z$ plane, the integration path is chosen as
        \begin{align}
            C_{c3}&=C_{c3}^{(1)}\cup C_{c3}^{(2)}\cup C_{c3}^{(3)},\\
            C_{c3}^{(1)}&=\Big\{z(t)\in \mathbb{C}\Big|z(t)=-1+(z_1+1)t, 0 \leq t \leq 1\Big\}, \\
            C_{c3}^{(2)}&=\Big\{z(t)\in \mathbb{C}\Big|z(t)=z_1+(z_2-z_1)t, 0 \leq t \leq 1\Big\}, \\
            C_{c3}^{(3)}&=\Big\{z(t)\in \mathbb{C}\Big|z(t)=z_2+(1-z_2)t, 0 \leq t \leq 1\Big\},
        \end{align} 
        like (c.3) in Fig.~\ref{fig:integrate_path}, where $z_1=\mathrm{Re}[z_{0\pi}]+i/10$, $z_2=\mathrm{Re}[z_{0R}]-i/10$.
        \item[(4)] $\mathrm{Im}[z_R]<0\;\& \;\mathrm{Im}[z_\pi]>0$. The two singularities are on  opposite sides of the real axis on the $z$ plane, and the integration path is chosen as
        \begin{align}
            C_{c4}&=C_{c4}^{(1)}\cup C_{c4}^{(2)}\cup C_{c4}^{(3)},\\
            C_{c4}^{(1)}&=\Big\{z(t)\in \mathbb{C}\Big|z(t)=-1+(z_3+1)t, 0 \leq t \leq 1\Big\}, \\
            C_{c4}^{(2)}&=\Big\{z(t)\in \mathbb{C}\Big|z(t)=z_3+(z_4-z_3)t, 0 \leq t \leq 1\Big\}, \\
            C_{c4}^{(3)}&=\Big\{z(t)\in \mathbb{C}\Big|z(t)=z_4+(1-z_4)t, 0 \leq t \leq 1\Big\},
        \end{align} 
        like (c.4) in Fig.~\ref{fig:integrate_path}, where $z_3=\mathrm{Re}[z_{0\pi}]-i/10$, $z_4=\mathrm{Re}[z_{0R}]+i/10$.
    \end{enumerate}
\end{enumerate}}

\subsection{\texorpdfstring{$D^*$}{D*} selfenergy}

The Green's function in Eq.\eqref{eq:LSE} is expressed as
    \begin{align}
        G(E;l)=\begin{pmatrix}
            G_0(E;l)& 0\\
            0   &G_{\pm}(E;l)
        \end{pmatrix}=\begin{pmatrix}
            \frac{1}{E-{l^2}/{(2\mu_0)}+{i}\Gamma_{0}^\mathcal{D}(E;l)/2 }& 0\\
              0 &\frac{1}{E-\Delta-{l^2}/{(2\mu_\pm)}+{i}{2}\Gamma_{\pm}^\mathcal{D}(E;l)/2}
        \end{pmatrix},
    \end{align}
    with $\mu_0={M_{D^0}M_{D^{\ast0}}}/{(M_{D^0}+M_{D^{\ast0}})}$ and $\mu_\pm={M_{D^+}M_{D^{\ast-}}}/{(M_{D^+}+M_{D^{\ast-}})}$. The energy-dependent decay widths of  neutral and charged $D^*$ mesons read
    \begin{align}
    \Gamma_{0}^{\mathcal{D}}(E; l)&=  \Gamma_{\left[D^{* 0} \rightarrow D^0 \gamma\right]}+\frac{g^2 M_{D^+}}{12 \pi F_\pi^2 M_{D^{*0}}}\left[\Sigma_{D^{+} \pi^{-} D^{0}}\left(E; l, \mu_\pm\right)-\Sigma_{D^{+} \pi^{-} D^{0}}\left(0; 0, \mu_\pm\right)\right]\nonumber\\
    &\quad +\frac{g^2 M_{D^0}}{24 \pi F_\pi^2 M_{D^{*0}}} 
    \Sigma_{D^0 \pi^0 D^{0}}\left(E; l, \mu_0\right) ,\\
    \Gamma_{\pm}^{\mathcal{D}}(E; l)&=  \Gamma_{\left[D^{*+} \rightarrow D^{+} \gamma\right]}
    +\frac{g^2 M_{D^+}}{24 \pi F_\pi^2 M_{D^{*+}}} \Sigma_{D^{+} \pi^0 D^-}\left(E; l, \mu_\pm\right)+\frac{g^2 M_{D^0}}{12 \pi F_\pi^2 M_{D^{*+}}} \Sigma_{D^0 \pi^{+} D^-}\left(E; l, \mu_\pm\right). 
    \end{align}
    We take the radiative decay widths as constants~\cite{ParticleDataGroup:2022pth},
    \begin{align}
        \Gamma_{[D^{\ast0}\to D^0\gamma]}=19.5~\text{keV},\quad\Gamma_{[D^{\ast+}\to D^+\gamma]}=1.3~\text{keV},
    \end{align}
    while the strong decay width depends on the pion momentum cubed in the rest framework of $D^*$,
    \begin{align}
        \Sigma_{\alpha \beta \zeta}(E; l, \mu)=\left[2 \mu_{\alpha \beta}\left(\sqrt{s}-M_{\alpha}-M_{\beta}-M_{\zeta}-\frac{l^2}{2 \mu}\right)\right]^{3 / 2},
        \label{eq:pion_momentum_cubic_in_width}
    \end{align}
  {with $\mu_{\alpha\beta}={M_{\alpha} M_{\beta}}/{(M_{\alpha}+M_{\beta})}.$ The cut of the square root function in  $\Sigma_{\alpha \beta\zeta}(E; l, \mu)$
is defined parallel to the negative imaginary axis on the complex energy plane to ensure a smooth cut crossing in the pole searching procedure~\cite{Doring:2009yv,Du:2021zzh,Ji:2022blw,Dong:2024fjk}.}
    
\subsection{Charged isospin-1 channel}

    The charged isospin-1 state of $D\bar D^*$, named as $W_{c1}^{\pm}$, can be expressed in flavor space as
    \begin{align}
       \left|W_{c1}^{+}\right\rangle = \frac1{\sqrt2}\left(\left|D^+\bar D^{*0}\right\rangle-\left|\bar D^0 D^{*+}\right\rangle\right),\quad 
       \left| W_{c1}^{-}\right\rangle = \frac1{\sqrt2}\left(\left|D^0D^{*-}\right\rangle-\left|D^- D^{*0}\right\rangle\right).
    \end{align}
    The pole positions of $W_{c1}^{\pm}$ are determined in a single-channel treatment by using the isospin averaged masses for the charmed mesons, $M_{D^{(*)}}=\left(M_{D^{(*)0}}+M_{D^{(*)+}}\right)/2$.
    The potential for the charged isospin-1 state of $D\bar D^*$ reads
    \begin{align}
    V_{+}(E_+;p^\prime,p)=C_{1X}-\frac{g^2}{12 F_{\pi}^2}V^{SS}_{D^+\pi^0\bar{D}^0}(E_+;p^\prime,p),
    \end{align}
    with $E_+=\sqrt{s}-M_{D}-M_{D^\ast}$. The OPE potential $V^{SS}_{D^+\pi^0\bar{D}^0}$ is defined as in Eq.~\eqref{eq:S_wave_projection}. The Green's function in this channel reads
    \begin{align}
       G_+(E_+;l)=\frac{1}{E_+-\frac{l^2}{2\mu_+}+\frac{i}{2}\Gamma_{+}^\mathcal{D}(E_+;l) },
    \end{align}
    with $\mu_+={M_{D}M_{D^{\ast}}}/{(M_{D}+M_{D^{\ast}})}$ and the energy-dependent decay width is taken to be the average of those of $D^{*0}$ and $D^{*+}$,
    \begin{align}
        \Gamma_{+}^\mathcal{D}(E_+;l)=\frac1{2}\left[{\Gamma_{0}^\mathcal{D}(E_+;l)+\Gamma_{\pm}^\mathcal{D}(E_+;l)}\right].
    \end{align}
    
\subsection{Riemann sheets}
\label{app:RSs}

For each channel, the scattering amplitude is in general a multi-valued function of the energy in the c.m. frame since it depends on the on-shell momentum, $l_{\rm on}$, determined by
\begin{align}
    \frac1{G_{0}(E;l_{\rm on,0})}=0,\ \frac1{G_{\pm}(E;l_{\rm on,\pm})}=0,\ \text{or}\ \frac1{G_{+}(E_+;l_{\rm on,+})}=0,
\end{align}
for a given $E$ or $E_+$. $l_{\rm on}$ has a cut from the threshold, which is on the complex plane because of the finite width of $D^*$, to infinity. We choose the cut to be a right-hand one.

For the neutral coupled channels, there are two channels and in turn 4 RSs, which are labeled with $(\mathrm{sgn}[\mathrm{Im}\,l_{\rm on,0}],\mathrm{sgn}[\mathrm{Im}\,l_{\rm on,\pm}])$,
\begin{align}
   \text{RS}_{++},\ \text{RS}_{-+},\ \text{RS}_{--},\ \text{RS}_{+-}.
\end{align}
The $\X$ pole, as a bound state pole, is located on $\text{RS}_{++}$ {and also has a shadow pole at $\text{RS}_{--}$}, while the $W_{c1}^0$ virtual state pole is located on {$\text{RS}_{+-}$}. For the charged single channel, there are 2 RSs labeled with $\mathrm{sgn}[\mathrm{Im}\,l_{\rm on,+}]$,
\begin{align}
\text{RS}_+ ,\ \text{RS}_-,
\end{align}
and the $W_{c1}^\pm$ virtual state pole is located on $\text{RS}_-$.

\subsection{Numerical procedure}

With all the singularities in the OPE potential and $D^{\ast}$ selfenergy being treated properly, the numerical solution of the integral LSE can be carried out. The integration over momentum $l$ in the numerical LSE is regularized by a sharp cutoff $\Lambda$. 
The momentum integration is performed using the Gauss-Legendre quadrature so that the integral equation becomes a matrix equation.
For each channel, the grid points are chosen such that they are much denser in the interval $l\in[0,\Lambda/10]$ than in $l\in[\Lambda/10,\Lambda]$ since the integrand varies more rapidly in the former small interval.
Note that the potential never becomes singular on the real-$l$ axis,
when $\Gamma_{{[D^\ast\to D\gamma]}}\neq 0$.
The determination of the $W_{c1}$ pole positions involves the following sequential steps:
\begin{enumerate}
    \item Determine the two LECs 
    $C_{0X}$ and $C_{1X}$ by solving the following two equations,
    \begin{align}
        & T^{-1}\left(-\mathcal{B}_X-i\dfrac{\Gamma_{X0}}{2}\right)=0, \\
        & R_{\pm/0} = \lim_{E\to -\mathcal{B}_X-i\Gamma_{X0}/2} \frac{T_{21}(E)}{T_{11}(E)},
    \end{align}
    where $\Gamma_{X0}$ is determined by the iterative procedure mentioned in the main text. The resulting LECs and $X(3872)$ pole position on RS$(+,+)$ are rechecked so that they indeed satisfy the above two equations.
    \item The LECs determined from the first step are used as inputs to solve $$T^{-1}\left(-\mathcal{B}_{W_{c1}^0}-i{\Gamma_{W_{c1}^0}}/{2}\right)=0$$ to obtain the pole position of $W_{c1}^0$ on RS$(-,-)$ in the neutral coupled channel system, and $C_{1X}$ is also used
    to solve $$T_{+}^{-1}\left(-\mathcal{B}_{W_{c1}^\pm}-i{\Gamma_{W_{c1}^\pm}}/{2}\right)=0$$ to obtain the $W_{c1}^\pm$ pole position on RS$(-)$ in the charged single channel system.
    \item For a given $\mathcal{B}_X$ in the range of $[0,180]$ keV, the last two steps are repeated by varying $R_X$ from 0.25 to 0.33 to assess the uncertainties. The outcomes obtained at $R_X=0.29$ are considered as the central values, while those at $R_X=0.25$ and $0.33$ are treated as the boundaries.
    \item The cutoff dependence of the $W_{c1}$ pole positions is checked numerically by varying $\Lambda$ from $0.5$ to $1$ GeV, and the pole positions in $E$ exhibit a minor change of less than $5\%$, which is indeed of $\mathcal{O}(Q^2/\Lambda^2)$.
\end{enumerate}

\section{Results in the pionless theory}
\label{app:nopion}

In the pionless theory, the LSE for the $(D\bar D^{*})_0^{[C=+]}${-}$(D\bar D^{*})_\pm^{[C=+]}$ coupled channels in Eq.~\eqref{eq:LSE} is reduced to
\begin{align}
T(E)=V_{\mathrm{ct}}+V_{\mathrm{ct}}J(E)T(E)=\frac1{V_{\mathrm{ct}}^{-1}-J(E)}
\end{align}
where 
\begin{align}
        J(E)=\begin{pmatrix}
            -\frac{\mu_0\Lambda}{\pi^2}-\frac{i\mu_0}{2\pi}k_0& 0\\
              0 &-\frac{\mu_\pm\Lambda}{\pi^2}-\frac{i\mu_\pm}{2\pi}k_\pm
        \end{pmatrix}\equiv\begin{pmatrix}
            d_{0}^\Lambda-\frac{i\mu_0}{2\pi}k_0& 0 \\
             0  &d_{\pm}^\Lambda-\frac{i\mu_\pm}{2\pi}k_\pm
        \end{pmatrix},
\end{align}
with 
\begin{align}
    k_0=\sqrt{2\mu_0E+i\mu_0\Gamma_{D^{\ast0}}}, \quad 
    k_\pm=\sqrt{2\mu_\pm(E-\Delta)+i\mu_\pm\Gamma_{D^{\ast+}}} ,
\end{align}
the on-shell c.m. momenta in $D^0\bar D^{*0}$ and $D^+D^{*-}$ channels, respectively. Here, $\Gamma_{D^{\ast0}}=55.3~\text{keV}$~\cite{Guo:2019qcn} and $\Gamma_{D^{\ast+}}=83.4~\text{keV}$~\cite{ParticleDataGroup:2022pth} are taken to be constants.

The cutoff dependence can be fully absorbed by $V_{\rm ct}$ when isospin breaking is neglected in the short-distance contributions, $d_{0}^\Lambda=
d_{\pm}^\Lambda \equiv d^\Lambda$,
and hence 
\begin{align}
T^{-1}(E)=V_{\mathrm{ct}}^{-1}-J(E)=(V_{\mathrm{ct}}^{R})^{-1}-J^R(E),
\end{align}
with
\begin{align}
V_{\mathrm{ct}}^R=\frac12\begin{pmatrix}[1.5]C_{0X}^R+C_{1X}^R&C_{0X}^R-C_{1X}^R\\C_{0X}^R-C_{1X}^R&C_{0X}^R+C_{1X}^R\end{pmatrix},
\quad
\text{and} 
\quad
        J^{R}(E)=\begin{pmatrix}
            -\frac{i\mu_0}{2\pi}k_0& 0 \\
              0 &-\frac{i\mu_\pm}{2\pi}k_\pm
        \end{pmatrix}.
\end{align}
The bare and renormalized LECs are related by
\begin{align}
    C_{0X}^{-1}=(C_{0X}^R)^{-1}+d^{\Lambda},\quad 
    C_{1X}^{-1}=(C_{1X}^R)^{-1}+d^{\Lambda}.
\end{align}
The LO inverse $T$ matrix can also be expressed in terms of the scattering lengths as
\begin{align}
    T^{-1}(E)=\begin{pmatrix}
       \frac{\mu_0}{2\pi}\left(-\frac{1}{a_{11}}+ik_0\right) &\frac{\sqrt{\mu_0\mu_{\pm}}}{2\pi}\frac{1}{a_{12}} \\ \frac{\sqrt{\mu_0\mu_{\pm}}}{2\pi}\frac{1}{a_{12}} &\frac{\mu_\pm}{2\pi}\left(-\frac{1}{a_{11}}+ik_\pm\right)
    \end{pmatrix},
\end{align}
where the scattering lengths can be expressed by the LECs as
 \begin{align}
     a_{11}=-\frac{\mu_0}{2\pi}\frac{C_{0X}^RC_{1X}^R}{C_{0X}^R+C_{1X}^R},\quad
     a_{12}=-\frac{\sqrt{\mu_0\mu_{\pm}}}{2\pi}\frac{C_{0X}^RC_{1X}^R}{C_{0X}^R-C_{1X}^R}.
 \end{align}

\begin{figure}[tbh]
    \centering
    \includegraphics[width=0.618\linewidth]{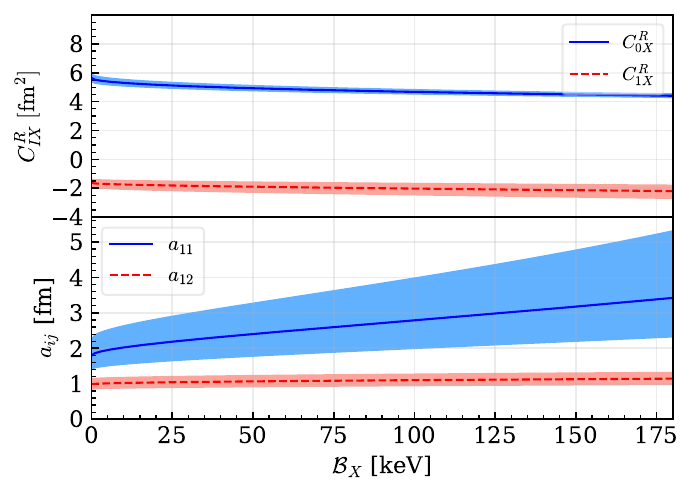}\hfill
    \includegraphics[width=0.618\linewidth]{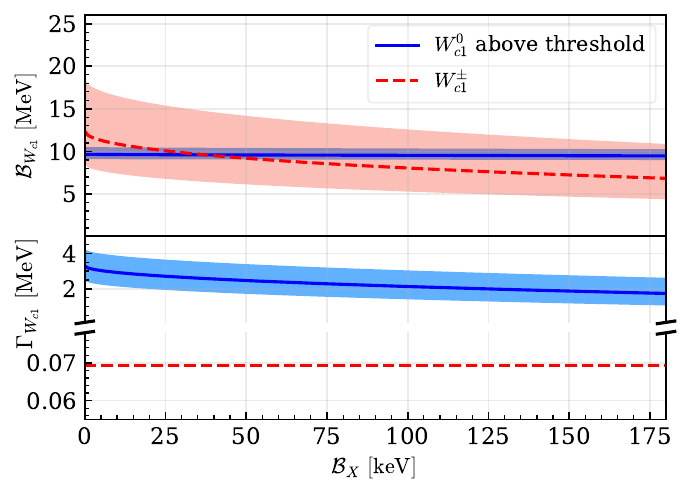}
    \caption{Left: Renormalized $C^R_{0X}$ and $C^R_{1X}$ and scattering lengths $a_{11}$ and $a_{12}$ as functions of $\mathcal{B}_X$. Upper: $C^R_{0X}$ and $C^R_{1X}$. Lower: scattering lengths $a_{11}$ and $a_{12}$. Right: Dependence of the predicted $W_{c1}$ virtual state pole position on the input $\X$ binding energy in the pionless theory. The bands are from the uncertainty in Eq.~\eqref{eq:Rpm0}. Upper: {magnitude} of the real part of the pole position with respect to the $D^0\bar D^{*0}$ (for $W_{c1}^0$, above threshold) or $D^0 D^{*-}$ (for $W_{c1}^-$, below threshold) threshold.
    Lower: twice the magnitude of the imaginary part of the $W_{c1}$ pole position. The bands are due to the uncertainty in Eq.~\eqref{eq:Rpm0}.}
    \label{fig:LECsnopi}
\end{figure}

\begin{figure}[tbh]
    \centering
    \includegraphics[width=0.618\linewidth]{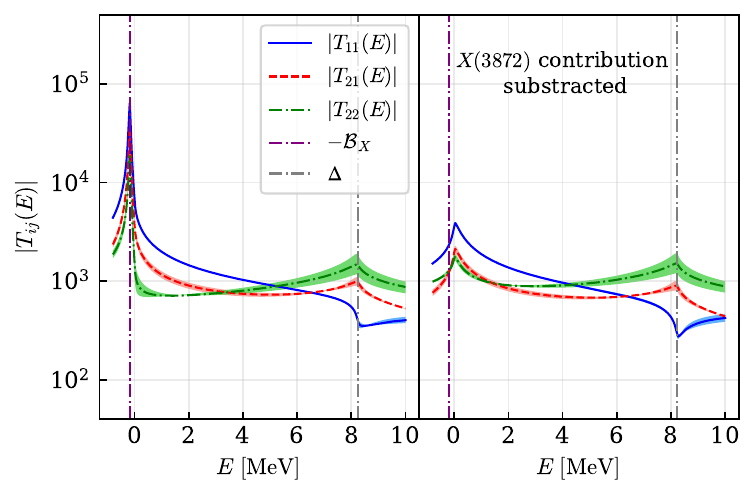}\hfill
    \includegraphics[width=0.618\linewidth]{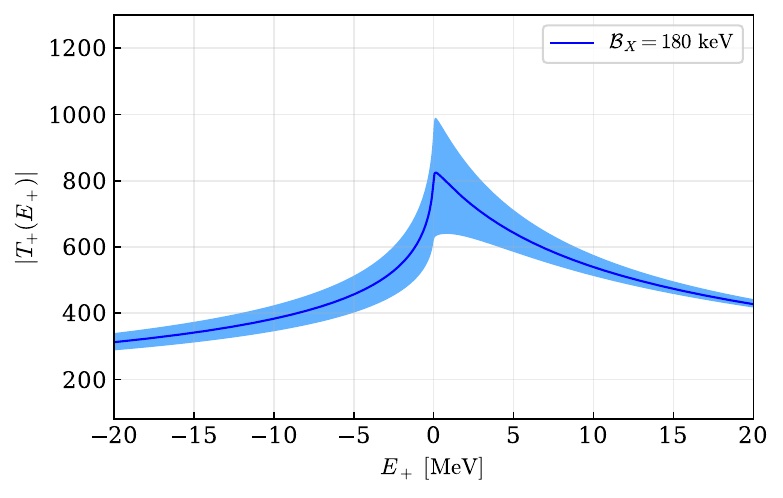}
    \caption{Left: Line shapes of the $S$-wave $D^0\bar D^{*0}$-$D^+D^{*-}$ scattering $T$-matrix elements in the pionless theory, where the left and right parts are for the full $T$-matrix elements and the $T$-matrix elements with the $\X$ pole contribution subtracted, respectively. 
    Right: Line shape of the single-channel $D^+\bar D^{*0}$ scattering $T$-matrix element, with $E_+$ defined relative to the $D^+\bar D^{*0}$ threshold.
    Here $\mathcal{B}_X$ is set to 180~keV and the uncertainty is from Eq.~\eqref{eq:Rpm0}. 
    }
    \label{fig:lineshape_fullnopi}
\end{figure}

Using the $X(3872)$ binding energy and the decay ratio $R_X$ as inputs, $C_{0X}^R$ and $C_{1X}^R$ can be solved as
\begin{align}
C_{0X}^{R}  = \frac{[1+R_{\pm/0}][J_{1}^R-R_{\pm/0}J_2^R]}{(J_1^R)^2-R_{\pm/0}^2(J_2^R)^2}|_{E = -\mathcal{B}_X},\quad 
C_{1X}^{R}  = \frac{[1-R_{\pm/0}][J_{1}^R+R_{\pm/0}J_2^R]}{(J_1^R)^2-R_{\pm/0}^2(J_2^R)^2}|_{E = -\mathcal{B}_X},
\end{align}
and as functions of $\mathcal{B}_X$ {as well as the scattering lengths}, shown in the left panel of Fig.~\ref{fig:LECsnopi}.
The values of $a_{11}$ and $a_{12}$ show that the situation corresponds to Case V4 in the classification of near-threshold coupled-channel line shapes in Ref.~\cite{Zhang:2024qkg}. From Table~II in Ref.~\cite{Zhang:2024qkg}, one can read off that the relevant poles should be on RS$_{++}$ for the $\X$ and RS$_{+-}$ for the $W_{c1}^0$.

For the charged isospin-1 single channel, the LSE reads
\begin{align}
T_+(E_+)=\frac{1}{C_{1X}^{-1}-J_+(E_+)} =\frac{1}{(C_{1X}^{R})^{-1}-J_+^R(E_+)},
\end{align}
where
\begin{align}
J_+=-\frac{\mu_+\Lambda}{\pi^2}-\frac{i\mu_+}{2\pi}k_+\equiv d^\Lambda_+-\frac{i\mu_+}{2\pi}k_+, \quad 
    J_+^R=    -\frac{i\mu_+}{2\pi}k_+,
\end{align}
with $k_+=\sqrt{2\mu_+E_++i\mu_+\Gamma_{D^{\ast}}}$,
and $\Gamma_{D^{\ast}}=(\Gamma_{D^{\ast0}}+\Gamma_{D^{\ast+}})/2$, and again short-distance isospin breaking is neglected such that $d_+^\Lambda=d^\Lambda$.

Using the determined $C_{0X}^R$ and $C_{1X}^R$, we predict the  $W_{c1}^0$ and $W_{c1}^\pm$ pole positions as functions of $\mathcal{B}_X$, shown in the right panel of Fig.~\ref{fig:LECsnopi}. Comparing them with the results with fully dynamical pions in Fig.~\ref{fig:Wc1_poles}, one sees that the real parts are similar while the imaginary parts differ significantly. This is expected as the imaginary parts are dominated by the $D\bar D\pi$ cuts.

Setting $\mathcal{B}_X=180$~keV, the predicted lineshape of the $(D\bar D^{*})_0^{[C=+]}${-}$(D\bar D^{*})_\pm^{[C=+]}$ coupled-channel $T$ matrix and the isovector $D^+\bar D^{*0}$ single-channel scattering $T$-matrix elements are illustrated in Fig.~\ref{fig:lineshape_fullnopi}. The contact term for the $J_I^{PC}=2^{++}_1$ $D^*\bar D^*$ system is also $C_{1X}$. Using the determined values of $C_{1X}^R$ in Fig.~\ref{fig:LECsnopi}, the $W_{c2}$ mass is predicted to be $4010.4^{+2.2}_{-3.5}$~MeV from the pole position of the $D^*\bar D^*$ single channel $T$ matrix, where the isospin-averaged mass of $D^*$ is taken.

\end{appendix}

\bibliographystyle{JHEP}
\bibliography{refs}

\providecommand{\noopsort}[1]{}

\providecommand{\href}[2]{#2}\begingroup\raggedright\begin{thebibliography}{10}

\bibitem{Hosaka:2016pey}
A.~Hosaka, T.~Iijima, K.~Miyabayashi, Y.~Sakai and S.~Yasui, \emph{Exotic
  hadrons with heavy flavors: {{X}}, {{Y}}, {{Z}}, and related states},
  \href{https://doi.org/10.1093/ptep/ptw045}{\emph{PTEP} {\bfseries 2016}
  (2016) 062C01} [\href{https://arxiv.org/abs/1603.09229}{{\ttfamily
  1603.09229}}].

\bibitem{Esposito:2016noz}
A.~Esposito, A.~Pilloni and A.D.~Polosa, \emph{Multiquark resonances},
  \href{https://doi.org/10.1016/j.physrep.2016.11.002}{\emph{Phys. Rept.}
  {\bfseries 668} (2017) 1} [\href{https://arxiv.org/abs/1611.07920}{{\ttfamily
  1611.07920}}].

\bibitem{Guo:2017jvc}
F.-K.~Guo, C.~Hanhart, U.-G.~Mei{\ss}ner, Q.~Wang, Q.~Zhao and B.-S.~Zou,
  \emph{Hadronic molecules},
  \href{https://doi.org/10.1103/RevModPhys.90.015004}{\emph{Rev. Mod. Phys.}
  {\bfseries 90} (2018) 015004}
  [\href{https://arxiv.org/abs/1705.00141}{{\ttfamily 1705.00141}}].

\bibitem{Olsen:2017bmm}
S.L.~Olsen, T.~Skwarnicki and D.~Zieminska, \emph{Nonstandard heavy mesons and
  baryons: {{Experimental}} evidence},
  \href{https://doi.org/10.1103/RevModPhys.90.015003}{\emph{Rev. Mod. Phys.}
  {\bfseries 90} (2018) 015003}
  [\href{https://arxiv.org/abs/1708.04012}{{\ttfamily 1708.04012}}].

\bibitem{Karliner:2017qhf}
M.~Karliner, J.L.~Rosner and T.~Skwarnicki, \emph{Multiquark states},
  \href{https://doi.org/10.1146/annurev-nucl-101917-020902}{\emph{Ann. Rev.
  Nucl. Part. Sci.} {\bfseries 68} (2018) 17}
  [\href{https://arxiv.org/abs/1711.10626}{{\ttfamily 1711.10626}}].

\bibitem{Brambilla:2019esw}
N.~Brambilla, S.~Eidelman, C.~Hanhart, A.~Nefediev, C.-P.~Shen, C.E.~Thomas
  et~al., \emph{The {{{\emph{XYZ}}}} states: {{Experimental}} and theoretical
  status and perspectives},
  \href{https://doi.org/10.1016/j.physrep.2020.05.001}{\emph{Phys. Rept.}
  {\bfseries 873} (2020) 1} [\href{https://arxiv.org/abs/1907.07583}{{\ttfamily
  1907.07583}}].

\bibitem{Yang:2020atz}
G.~Yang, J.~Ping and J.~Segovia, \emph{Tetra- and penta-quark structures in the
  constituent quark model},
  \href{https://doi.org/10.3390/sym12111869}{\emph{Symmetry} {\bfseries 12}
  (2020) 1869} [\href{https://arxiv.org/abs/2009.00238}{{\ttfamily
  2009.00238}}].

\bibitem{Chen:2022asf}
H.-X.~Chen, W.~Chen, X.~Liu, Y.-R.~Liu and S.-L.~Zhu, \emph{An updated review
  of the new hadron states},
  \href{https://doi.org/10.1088/1361-6633/aca3b6}{\emph{Rept. Prog. Phys.}
  {\bfseries 86} (2022) 026201}
  [\href{https://arxiv.org/abs/2204.02649}{{\ttfamily 2204.02649}}].

\bibitem{Meng:2022ozq}
L.~Meng, B.~Wang, G.-J.~Wang and S.-L.~Zhu, \emph{Chiral perturbation theory
  for heavy hadrons and chiral effective field theory for heavy hadronic
  molecules}, \href{https://doi.org/10.1016/j.physrep.2023.04.003}{\emph{Phys.
  Rept.} {\bfseries 1019} (2023) 2266}
  [\href{https://arxiv.org/abs/2204.08716}{{\ttfamily 2204.08716}}].

\bibitem{ParticleDataGroup:2022pth}
{\scshape Particle Data Group} collaboration, \emph{Review of {{Particle
  Physics}}}, \href{https://doi.org/10.1093/ptep/ptac097}{\emph{PTEP}
  {\bfseries 2022} (2022) 083C01}.

\bibitem{Belle:2003nnu}
{\scshape Belle} collaboration, \emph{Observation of a narrow charmonium-like
  state in exclusive
  {{B}}{\textsuperscript{{\textpm}}}{$\rightarrow$}{{K}}{\textsuperscript{{\textpm}}}{$\pi^{+}\pi^{-}$}{{J}}/{$\psi$}
  decays}, \href{https://doi.org/10.1103/PhysRevLett.91.262001}{\emph{Phys.
  Rev. Lett.} {\bfseries 91} (2003) 262001}
  [\href{https://arxiv.org/abs/hep-ex/0309032}{{\ttfamily hep-ex/0309032}}].

\bibitem{D0:2004zmu}
{\scshape D0} collaboration, \emph{{Observation and properties of the $X(3872)$
  decaying to $J/\psi \pi^+ \pi^-$ in $p\bar{p}$ collisions at $\sqrt{s} =
  1.96$ TeV}}, \href{https://doi.org/10.1103/PhysRevLett.93.162002}{\emph{Phys.
  Rev. Lett.} {\bfseries 93} (2004) 162002}
  [\href{https://arxiv.org/abs/hep-ex/0405004}{{\ttfamily hep-ex/0405004}}].

\bibitem{Belle:2006olv}
{\scshape Belle} collaboration, \emph{{Observation of a near-threshold $D^0
  \bar{D}^0 \pi^0$ enhancement in $B \rightarrow D^0 \bar{D}^0 \pi^0 K$
  decay}}, \href{https://doi.org/10.1103/PhysRevLett.97.162002}{\emph{Phys.
  Rev. Lett.} {\bfseries 97} (2006) 162002}
  [\href{https://arxiv.org/abs/hep-ex/0606055}{{\ttfamily hep-ex/0606055}}].

\bibitem{BaBar:2010wfc}
{\scshape BaBar} collaboration, \emph{Evidence for the decay {{$X(3872)
  \rightarrow J/\psi \omega$}}},
  \href{https://doi.org/10.1103/PhysRevD.82.011101}{\emph{Phys. Rev. D}
  {\bfseries 82} (2010) 011101}
  [\href{https://arxiv.org/abs/1005.5190}{{\ttfamily 1005.5190}}].

\bibitem{LHCb:2013kgk}
{\scshape LHCb} collaboration, \emph{Determination of the {{X}}(3872) meson
  quantum numbers},
  \href{https://doi.org/10.1103/PhysRevLett.110.222001}{\emph{Phys. Rev. Lett.}
  {\bfseries 110} (2013) 222001}
  [\href{https://arxiv.org/abs/1302.6269}{{\ttfamily 1302.6269}}].

\bibitem{BESIII:2013fnz}
{\scshape BESIII} collaboration, \emph{Observation of {{$e^+e^- \rightarrow
  \gamma X(3872)$}} at {{BESIII}}},
  \href{https://doi.org/10.1103/PhysRevLett.112.092001}{\emph{Phys. Rev. Lett.}
  {\bfseries 112} (2014) 092001}
  [\href{https://arxiv.org/abs/1310.4101}{{\ttfamily 1310.4101}}].

\bibitem{ATLAS:2016kwu}
{\scshape ATLAS} collaboration, \emph{Measurements of {{$\psi$(2S)}} and
  {{X(3872) $\rightarrow$J/$\psi\pi^{+}\pi^{-}$}} production in {{pp}}
  collisions at {{s = 8}} {{TeV}} with the {{ATLAS}} detector},
  \href{https://doi.org/10.1007/JHEP01(2017)117}{\emph{JHEP} {\bfseries 01}
  (2017) 117} [\href{https://arxiv.org/abs/1610.09303}{{\ttfamily
  1610.09303}}].

\bibitem{BESIII:2019esk}
{\scshape BESIII} collaboration, \emph{Observation of the decay {{$X(3872)
  \rightarrow\pi^0\chi_{c1}(1P)$}}},
  \href{https://doi.org/10.1103/PhysRevLett.122.202001}{\emph{Phys. Rev. Lett.}
  {\bfseries 122} (2019) 202001}
  [\href{https://arxiv.org/abs/1901.03992}{{\ttfamily 1901.03992}}].

\bibitem{CMS:2021znk}
{\scshape CMS} collaboration, \emph{{Evidence for X(3872) in Pb-Pb Collisions
  and Studies of its Prompt Production at $\sqrt {s_{NN}}$=5.02\,\,TeV}},
  \href{https://doi.org/10.1103/PhysRevLett.128.032001}{\emph{Phys. Rev. Lett.}
  {\bfseries 128} (2022) 032001}
  [\href{https://arxiv.org/abs/2102.13048}{{\ttfamily 2102.13048}}].

\bibitem{BESIII:2022bse}
{\scshape BESIII} collaboration, \emph{Observation of a new {{$X$}}(3872)
  production process {{$e^+e^{-}\rightarrow\omega X(3872)$}}},
  \href{https://doi.org/10.1103/PhysRevLett.130.151904}{\emph{Phys. Rev. Lett.}
  {\bfseries 130} (2023) 151904}
  [\href{https://arxiv.org/abs/2212.07291}{{\ttfamily 2212.07291}}].

\bibitem{LHCb:2020xds}
{\scshape LHCb} collaboration, \emph{Study of the lineshape of the
  {{$\chi${\textsubscript{c1}}(3872)}} state},
  \href{https://doi.org/10.1103/PhysRevD.102.092005}{\emph{Phys. Rev. D}
  {\bfseries 102} (2020) 092005}
  [\href{https://arxiv.org/abs/2005.13419}{{\ttfamily 2005.13419}}].

\bibitem{Belle:2008fma}
{\scshape Belle} collaboration, \emph{Study of the {$B\to X(3872)(\to
  D^{*0}\bar D^0)K$} decay},
  \href{https://doi.org/10.1103/PhysRevD.81.031103}{\emph{Phys. Rev. D}
  {\bfseries 81} (2010) 031103}
  [\href{https://arxiv.org/abs/0810.0358}{{\ttfamily 0810.0358}}].

\bibitem{Li:2019kpj}
C.~Li and C.-Z.~Yuan, \emph{Determination of the absolute branching fractions
  of {{X}}(3872) decays},
  \href{https://doi.org/10.1103/PhysRevD.100.094003}{\emph{Phys. Rev. D}
  {\bfseries 100} (2019) 094003}
  [\href{https://arxiv.org/abs/1907.09149}{{\ttfamily 1907.09149}}].

\bibitem{Braaten:2019ags}
E.~Braaten, L.-P.~He and K.~Ingles, \emph{Branching {{Fractions}} of the
  {{X}}(3872)}, {\emph{arXiv:1908.02807 [hep-ph]} (2019) }
  [\href{https://arxiv.org/abs/1908.02807}{{\ttfamily 1908.02807}}].

\bibitem{Close:2003sg}
F.E.~Close and P.R.~Page, \emph{The {{D}}{\textsuperscript{*0}}{{{\=D}}}{$^0$}
  threshold resonance},
  \href{https://doi.org/10.1016/j.physletb.2003.10.032}{\emph{Phys. Lett. B}
  {\bfseries 578} (2004) 119}
  [\href{https://arxiv.org/abs/hep-ph/0309253}{{\ttfamily hep-ph/0309253}}].

\bibitem{Pakvasa:2003ea}
S.~Pakvasa and M.~Suzuki, \emph{On the hidden charm state at 3872 {{MeV}}},
  \href{https://doi.org/10.1016/j.physletb.2003.11.005}{\emph{Phys. Lett. B}
  {\bfseries 579} (2004) 67}
  [\href{https://arxiv.org/abs/hep-ph/0309294}{{\ttfamily hep-ph/0309294}}].

\bibitem{Voloshin:2003nt}
M.~Voloshin, \emph{Interference and binding effects in decays of possible
  molecular component of {{X}}(3872)},
  \href{https://doi.org/10.1016/j.physletb.2003.11.014}{\emph{Phys. Lett. B}
  {\bfseries 579} (2004) 316}
  [\href{https://arxiv.org/abs/hep-ph/0309307}{{\ttfamily hep-ph/0309307}}].

\bibitem{Swanson:2003tb}
E.S.~Swanson, \emph{Short range structure in the {{X}}(3872)},
  \href{https://doi.org/10.1016/j.physletb.2004.03.033}{\emph{Phys. Lett. B}
  {\bfseries 588} (2004) 189}
  [\href{https://arxiv.org/abs/hep-ph/0311229}{{\ttfamily hep-ph/0311229}}].

\bibitem{Braaten:2003he}
E.~Braaten and M.~Kusunoki, \emph{Low-energy universality and the new
  charmonium resonance at 3870 {{MeV}}},
  \href{https://doi.org/10.1103/PhysRevD.69.074005}{\emph{Phys. Rev. D}
  {\bfseries 69} (2004) 074005}
  [\href{https://arxiv.org/abs/hep-ph/0311147}{{\ttfamily hep-ph/0311147}}].

\bibitem{Tornqvist:2004qy}
N.A.~T{\"o}rnqvist, \emph{Isospin breaking of the narrow charmonium state of
  {{Belle}} at 3872 {{MeV}} as a deuson},
  \href{https://doi.org/10.1016/j.physletb.2004.03.077}{\emph{Phys. Lett. B}
  {\bfseries 590} (2004) 209}
  [\href{https://arxiv.org/abs/hep-ph/0402237}{{\ttfamily hep-ph/0402237}}].

\bibitem{Kalashnikova:2018vkv}
Y.S.~Kalashnikova and A.V.~Nefediev, \emph{X(3872) in the molecular model},
  \href{https://doi.org/10.3367/UFNe.2018.08.038411}{\emph{Phys. Usp.}
  {\bfseries 62} (2019) 568}
  [\href{https://arxiv.org/abs/1811.01324}{{\ttfamily 1811.01324}}].

\bibitem{Maiani:2004vq}
L.~Maiani, F.~Piccinini, A.D.~Polosa and V.~Riquer, \emph{Diquark-antidiquarks
  with hidden or open charm and the nature of {{X}}(3872)},
  \href{https://doi.org/10.1103/PhysRevD.71.014028}{\emph{Phys. Rev. D}
  {\bfseries 71} (2005) 014028}
  [\href{https://arxiv.org/abs/hep-ph/0412098}{{\ttfamily hep-ph/0412098}}].

\bibitem{Belle:2011vlx}
{\scshape Belle} collaboration, \emph{Bounds on the width, mass difference and
  other properties of {{X(3872) $\rightarrow\pi^+$ $\pi^-$ J/$\psi$}} decays},
  \href{https://doi.org/10.1103/PhysRevD.84.052004}{\emph{Phys. Rev. D}
  {\bfseries 84} (2011) 052004}
  [\href{https://arxiv.org/abs/1107.0163}{{\ttfamily 1107.0163}}].

\bibitem{Matuschek:2020gqe}
I.~Matuschek, V.~Baru, F.-K.~Guo and C.~Hanhart, \emph{{On the nature of
  near-threshold bound and virtual states}},
  \href{https://doi.org/10.1140/epja/s10050-021-00413-y}{\emph{Eur. Phys. J. A}
  {\bfseries 57} (2021) 101}
  [\href{https://arxiv.org/abs/2007.05329}{{\ttfamily 2007.05329}}].

\bibitem{Isgur:1989vq}
N.~Isgur and M.B.~Wise, \emph{Weak decays of heavy mesons in the static quark
  approximation},
  \href{https://doi.org/10.1016/0370-2693(89)90566-2}{\emph{Phys. Lett. B}
  {\bfseries 232} (1989) 113}.

\bibitem{Nieves:2012tt}
J.~Nieves and M.P.~Valderrama, \emph{The {{Heavy Quark Spin Symmetry Partners}}
  of the {{X}}(3872)},
  \href{https://doi.org/10.1103/PhysRevD.86.056004}{\emph{Phys. Rev. D}
  {\bfseries 86} (2012) 056004}
  [\href{https://arxiv.org/abs/1204.2790}{{\ttfamily 1204.2790}}].

\bibitem{Hidalgo-Duque:2013pva}
C.~{Hidalgo-Duque}, J.~Nieves, A.~Ozpineci and V.~Zamiralov, \emph{X(3872) and
  its {{Partners}} in the {{Heavy Quark Limit}} of {{QCD}}},
  \href{https://doi.org/10.1016/j.physletb.2013.10.056}{\emph{Phys. Lett. B}
  {\bfseries 727} (2013) 432}
  [\href{https://arxiv.org/abs/1305.4487}{{\ttfamily 1305.4487}}].

\bibitem{Baru:2016iwj}
V.~Baru, E.~Epelbaum, A.A.~Filin, C.~Hanhart, U.-G.~Mei{\ss}ner and
  A.V.~Nefediev, \emph{Heavy-quark spin symmetry partners of the {{X}} (3872)
  revisited}, \href{https://doi.org/10.1016/j.physletb.2016.10.008}{\emph{Phys.
  Lett. B} {\bfseries 763} (2016) 20}
  [\href{https://arxiv.org/abs/1605.09649}{{\ttfamily 1605.09649}}].

\bibitem{Hu:2024hex}
T.-R.~Hu, S.~Chen and F.-K.~Guo, \emph{{Entanglement suppression and low-energy
  scattering of heavy mesons}},
  \href{https://doi.org/10.1103/PhysRevD.110.014001}{\emph{Phys. Rev. D}
  {\bfseries 110} (2024) 014001}
  [\href{https://arxiv.org/abs/2404.05958}{{\ttfamily 2404.05958}}].

\bibitem{Hidalgo-Duque:2012rqv}
C.~{Hidalgo-Duque}, J.~Nieves and M.P.~Valderrama, \emph{Light flavor and heavy
  quark spin symmetry in heavy meson molecules},
  \href{https://doi.org/10.1103/PhysRevD.87.076006}{\emph{Phys. Rev. D}
  {\bfseries 87} (2013) 076006}
  [\href{https://arxiv.org/abs/1210.5431}{{\ttfamily 1210.5431}}].

\bibitem{Guo:2013sya}
F.-K.~Guo, C.~{Hidalgo-Duque}, J.~Nieves and M.P.~Valderrama,
  \emph{Consequences of heavy quark symmetries for hadronic molecules},
  \href{https://doi.org/10.1103/PhysRevD.88.054007}{\emph{Phys. Rev. D}
  {\bfseries 88} (2013) 054007}
  [\href{https://arxiv.org/abs/1303.6608}{{\ttfamily 1303.6608}}].

\bibitem{Albaladejo:2015dsa}
M.~Albaladejo, F.-K.~Guo, C.~{Hidalgo-Duque}, J.~Nieves and M.P.~Valderrama,
  \emph{Decay widths of the spin-2 partners of the {{X}}(3872)},
  \href{https://doi.org/10.1140/epjc/s10052-015-3753-6}{\emph{Eur. Phys. J. C}
  {\bfseries 75} (2015) 547}
  [\href{https://arxiv.org/abs/1504.00861}{{\ttfamily 1504.00861}}].

\bibitem{Gamermann:2009fv}
D.~Gamermann and E.~Oset, \emph{Isospin breaking effects in the {{X}}(3872)
  resonance}, \href{https://doi.org/10.1103/PhysRevD.80.014003}{\emph{Phys.
  Rev. D} {\bfseries 80} (2009) 014003}
  [\href{https://arxiv.org/abs/0905.0402}{{\ttfamily 0905.0402}}].

\bibitem{Gamermann:2009uq}
D.~Gamermann, J.~Nieves, E.~Oset and E.R.~Arriola, \emph{Couplings in coupled
  channels versus wave functions: {{Application}} to the {{X}}(3872)
  resonance}, \href{https://doi.org/10.1103/PhysRevD.81.014029}{\emph{Phys.
  Rev. D} {\bfseries 81} (2010) 014029}.

\bibitem{Li:2012cs}
N.~Li and S.-L.~Zhu, \emph{Isospin breaking, coupled-channel effects, and
  {{X}}(3872)}, \href{https://doi.org/10.1103/PhysRevD.86.074022}{\emph{Phys.
  Rev. D} {\bfseries 86} (2012) 074022}
  [\href{https://arxiv.org/abs/1207.3954}{{\ttfamily 1207.3954}}].

\bibitem{Meng:2021kmi}
L.~Meng, G.-J.~Wang, B.~Wang and S.-L.~Zhu, \emph{Revisit the isospin violating
  decays of {{X}}(3872)},
  \href{https://doi.org/10.1103/PhysRevD.104.094003}{\emph{Phys. Rev. D}
  {\bfseries 104} (2021) 094003}
  [\href{https://arxiv.org/abs/2109.01333}{{\ttfamily 2109.01333}}].

\bibitem{Ji:2022uie}
T.~Ji, X.-K.~Dong, M.~Albaladejo, M.-L.~Du, F.-K.~Guo and J.~Nieves,
  \emph{Establishing the heavy quark spin and light flavor molecular multiplets
  of the {{X}}(3872), {{Z}}{\textsubscript{c}}(3900) and {{X}}(3960)},
  \href{https://doi.org/10.1103/PhysRevD.106.094002}{\emph{Phys. Rev. D}
  {\bfseries 106} (2022) 094002}
  [\href{https://arxiv.org/abs/2207.08563}{{\ttfamily 2207.08563}}].

\bibitem{Suzuki:2005ha}
M.~Suzuki, \emph{X(3872) boson: {{Molecule}} or charmonium},
  \href{https://doi.org/10.1103/PhysRevD.72.114013}{\emph{Phys. Rev. D}
  {\bfseries 72} (2005) 114013}
  [\href{https://arxiv.org/abs/hep-ph/0508258}{{\ttfamily hep-ph/0508258}}].

\bibitem{Hanhart:2011tn}
C.~Hanhart, Y.S.~Kalashnikova, A.E.~Kudryavtsev and A.V.~Nefediev,
  \emph{Remarks on the quantum numbers of {{X}}(3872) from the invariant mass
  distributions of the {{$\rho$J}}/{$\psi$} and {{$\omega$J}}/{$\psi$} final
  states}, \href{https://doi.org/10.1103/PhysRevD.85.011501}{\emph{Phys. Rev.
  D} {\bfseries 85} (2012) 011501}
  [\href{https://arxiv.org/abs/1111.6241}{{\ttfamily 1111.6241}}].

\bibitem{LHCb:2022jez}
{\scshape LHCb} collaboration, \emph{Observation of sizeable {$\omega$}
  contribution to
  {$\chi$}{\textsubscript{c1}}(3872){$\rightarrow\pi^{+}\pi^{-}$}{{J}}/{$\psi$}
  decays}, \href{https://doi.org/10.1103/PhysRevD.108.L011103}{\emph{Phys. Rev.
  D} {\bfseries 108} (2023) L011103}
  [\href{https://arxiv.org/abs/2204.12597}{{\ttfamily 2204.12597}}].

\bibitem{Braaten:2005jj}
E.~Braaten and M.~Kusunoki, \emph{Factorization in the production and decay of
  the {{X}}(3872)},
  \href{https://doi.org/10.1103/PhysRevD.72.014012}{\emph{Phys. Rev. D}
  {\bfseries 72} (2005) 014012}
  [\href{https://arxiv.org/abs/hep-ph/0506087}{{\ttfamily hep-ph/0506087}}].

\bibitem{Epelbaum:2008ga}
E.~Epelbaum, H.-W.~Hammer and U.-G.~Mei{\ss}ner, \emph{Modern theory of nuclear
  forces}, \href{https://doi.org/10.1103/RevModPhys.81.1773}{\emph{Rev. Mod.
  Phys.} {\bfseries 81} (2009) 1773}
  [\href{https://arxiv.org/abs/0811.1338}{{\ttfamily 0811.1338}}].

\bibitem{Wise:1992hn}
M.B.~Wise, \emph{{Chiral perturbation theory for hadrons containing a heavy
  quark}}, \href{https://doi.org/10.1103/PhysRevD.45.R2188}{\emph{Phys. Rev. D}
  {\bfseries 45} (1992) R2188}.

\bibitem{Zhang:2020mpi}
Z.-H.~Zhang and F.-K.~Guo, \emph{{$D^{\pm}D^{*\mp}$ Hadronic Atom as a Key to
  Revealing the $X(3872)$ Mystery}},
  \href{https://doi.org/10.1103/PhysRevLett.127.012002}{\emph{Phys. Rev. Lett.}
  {\bfseries 127} (2021) 012002}
  [\href{https://arxiv.org/abs/2012.08281}{{\ttfamily 2012.08281}}].

\bibitem{Dong:2024fjk}
X.-K.~Dong, T.~Ji, F.-K.~Guo, U.-G.~Mei{\ss}ner and B.-S.~Zou, \emph{Hints of
  the {$J^{PC}=0^{--}$} and {$1^{--}$} {$K^*\bar K_1(1270)$} molecules in the
  {{$J$}}/{$\psi\rightarrow\phi\eta\eta$}' decay}, {\emph{arXiv:2402.02903
  [hep-ph]} (2024) } [\href{https://arxiv.org/abs/2402.02903}{{\ttfamily
  2402.02903}}].

\bibitem{Doring:2009yv}
M.~D{\"o}ring, C.~Hanhart, F.~Huang, S.~Krewald and U.-G.~Meissner,
  \emph{{Analytic properties of the scattering amplitude and resonances
  parameters in a meson exchange model}},
  \href{https://doi.org/10.1016/j.nuclphysa.2009.08.010}{\emph{Nucl. Phys. A}
  {\bfseries 829} (2009) 170}
  [\href{https://arxiv.org/abs/0903.4337}{{\ttfamily 0903.4337}}].

\bibitem{Baru:2011rs}
V.~Baru, A.A.~Filin, C.~Hanhart, Y.S.~Kalashnikova, A.E.~Kudryavtsev and
  A.V.~Nefediev, \emph{Three-body {{D{\=D}$\pi$}} dynamics for the
  {{X}}(3872)}, \href{https://doi.org/10.1103/PhysRevD.84.074029}{\emph{Phys.
  Rev. D} {\bfseries 84} (2011) 074029}
  [\href{https://arxiv.org/abs/1108.5644}{{\ttfamily 1108.5644}}].

\bibitem{Du:2021zzh}
M.-L.~Du, V.~Baru, X.-K.~Dong, A.~Filin, F.-K.~Guo, C.~Hanhart et~al.,
  \emph{Coupled-channel approach to {{T}}{\textsubscript{cc}}{$^+$} including
  three-body effects},
  \href{https://doi.org/10.1103/PhysRevD.105.014024}{\emph{Phys. Rev. D}
  {\bfseries 105} (2022) 014024}
  [\href{https://arxiv.org/abs/2110.13765}{{\ttfamily 2110.13765}}].

\bibitem{Ji:2022blw}
T.~Ji, X.-K.~Dong, F.-K.~Guo and B.-S.~Zou, \emph{{Prediction of a Narrow
  Exotic Hadronic State with Quantum Numbers $J^{PC}=0^{--}$}},
  \href{https://doi.org/10.1103/PhysRevLett.129.102002}{\emph{Phys. Rev. Lett.}
  {\bfseries 129} (2022) 102002}
  [\href{https://arxiv.org/abs/2205.10994}{{\ttfamily 2205.10994}}].

\bibitem{Guo:2019qcn}
F.-K.~Guo, \emph{{Novel Method for Precisely Measuring the $X(3872)$ Mass}},
  \href{https://doi.org/10.1103/PhysRevLett.122.202002}{\emph{Phys. Rev. Lett.}
  {\bfseries 122} (2019) 202002}
  [\href{https://arxiv.org/abs/1902.11221}{{\ttfamily 1902.11221}}].

\bibitem{Zhang:2024qkg}
Z.-H.~Zhang and F.-K.~Guo, \emph{{Classification of Coupled-Channel
  Near-Threshold Structures}},
  \href{https://arxiv.org/abs/2407.10620}{{\ttfamily 2407.10620}}.

\bibitem{Eden:1963zz}
R.J.~Eden and J.R.~Taylor, \emph{{Resonance Multiplets and Broken Symmetry}},
  \href{https://doi.org/10.1103/PhysRevLett.11.516}{\emph{Phys. Rev. Lett.}
  {\bfseries 11} (1963) 516}.

\bibitem{Dong:2020hxe}
X.-K.~Dong, F.-K.~Guo and B.-S.~Zou, \emph{Explaining the many threshold
  structures in the heavy-quark hadron spectrum},
  \href{https://doi.org/10.1103/PhysRevLett.126.152001}{\emph{Phys. Rev. Lett.}
  {\bfseries 126} (2021) 152001}
  [\href{https://arxiv.org/abs/2011.14517}{{\ttfamily 2011.14517}}].

\bibitem{Baru:2024ptl}
V.~Baru, F.-K.~Guo, C.~Hanhart and A.~Nefediev, \emph{{How does the X(3872)
  show up in e+e- collisions: Dip versus peak}},
  \href{https://doi.org/10.1103/PhysRevD.109.L111501}{\emph{Phys. Rev. D}
  {\bfseries 109} (2024) L111501}
  [\href{https://arxiv.org/abs/2404.12003}{{\ttfamily 2404.12003}}].

\bibitem{BESIII:2022ner}
{\scshape BESIII} collaboration, \emph{Measurement of the
  {{e$^+$e$^{-}\rightarrow\pi^{+}\pi^{-}$J/$\psi$}} cross section in the
  vicinity of 3.872 {{GeV}}},
  \href{https://doi.org/10.1103/PhysRevD.107.032007}{\emph{Phys. Rev. D}
  {\bfseries 107} (2023) 032007}
  [\href{https://arxiv.org/abs/2209.12007}{{\ttfamily 2209.12007}}].

\bibitem{BESIII:2013ris}
{\scshape BESIII} collaboration, \emph{{Observation of a Charged Charmoniumlike
  Structure in $e^+e^- \to \pi^+\pi^- J/\psi$ at $\sqrt{s}$ =4.26 GeV}},
  \href{https://doi.org/10.1103/PhysRevLett.110.252001}{\emph{Phys. Rev. Lett.}
  {\bfseries 110} (2013) 252001}
  [\href{https://arxiv.org/abs/1303.5949}{{\ttfamily 1303.5949}}].

\bibitem{Belle:2013yex}
{\scshape Belle} collaboration, \emph{{Study of $e^+e^- \to \pi^+ \pi^- J/\pi$
  and Observation of a Charged Charmoniumlike State at Belle}},
  \href{https://doi.org/10.1103/PhysRevLett.110.252002}{\emph{Phys. Rev. Lett.}
  {\bfseries 110} (2013) 252002}
  [\href{https://arxiv.org/abs/1304.0121}{{\ttfamily 1304.0121}}].

\bibitem{BESIII:2013ouc}
{\scshape BESIII} collaboration, \emph{{Observation of a Charged Charmoniumlike
  Structure $Z_c$(4020) and Search for the $Z_c$(3900) in $e^+e^- \to
  \pi^+\pi^-h_c$}},
  \href{https://doi.org/10.1103/PhysRevLett.111.242001}{\emph{Phys. Rev. Lett.}
  {\bfseries 111} (2013) 242001}
  [\href{https://arxiv.org/abs/1309.1896}{{\ttfamily 1309.1896}}].

\bibitem{BESIII:2013mhi}
{\scshape BESIII} collaboration, \emph{{Observation of a charged charmoniumlike
  structure in $e^+e^- \to (D^{*} \bar{D}^{*})^{\pm} \pi^\mp$ at
  $\sqrt{s}=4.26$GeV}},
  \href{https://doi.org/10.1103/PhysRevLett.112.132001}{\emph{Phys. Rev. Lett.}
  {\bfseries 112} (2014) 132001}
  [\href{https://arxiv.org/abs/1308.2760}{{\ttfamily 1308.2760}}].

\bibitem{Wang:2018jlv}
Q.~Wang, V.~Baru, A.A.~Filin, C.~Hanhart, A.V.~Nefediev and J.L.~Wynen,
  \emph{{Line shapes of the $Z_b(10610)$ and $Z_b(10650)$ in the elastic and
  inelastic channels revisited}},
  \href{https://doi.org/10.1103/PhysRevD.98.074023}{\emph{Phys. Rev. D}
  {\bfseries 98} (2018) 074023}
  [\href{https://arxiv.org/abs/1805.07453}{{\ttfamily 1805.07453}}].

\bibitem{Baru:2019xnh}
V.~Baru, E.~Epelbaum, A.A.~Filin, C.~Hanhart, A.V.~Nefediev and Q.~Wang,
  \emph{{Spin partners $W_{bJ}$ from the line shapes of the $Z_b(10610)$ and
  $Z_b(10650)$}}, \href{https://doi.org/10.1103/PhysRevD.99.094013}{\emph{Phys.
  Rev. D} {\bfseries 99} (2019) 094013}
  [\href{https://arxiv.org/abs/1901.10319}{{\ttfamily 1901.10319}}].

\bibitem{BESIII:2020qkh}
{\scshape BESIII} collaboration, \emph{{Observation of a Near-Threshold
  Structure in the $K^+$ Recoil-Mass Spectra in $e^+e^- \rightarrow
  K^+(D_s^-D^{*0}+D_s^{*-}D^0$)}},
  \href{https://doi.org/10.1103/PhysRevLett.126.102001}{\emph{Phys. Rev. Lett.}
  {\bfseries 126} (2021) 102001}
  [\href{https://arxiv.org/abs/2011.07855}{{\ttfamily 2011.07855}}].

\bibitem{LHCb:2021uow}
{\scshape LHCb} collaboration, \emph{{Observation of New Resonances Decaying to
  $J/\psi K^+$+ and $J/\psi \phi$}},
  \href{https://doi.org/10.1103/PhysRevLett.127.082001}{\emph{Phys. Rev. Lett.}
  {\bfseries 127} (2021) 082001}
  [\href{https://arxiv.org/abs/2103.01803}{{\ttfamily 2103.01803}}].

\bibitem{Yang:2020nrt}
Z.~Yang, X.~Cao, F.-K.~Guo, J.~Nieves and M.P.~Valderrama, \emph{{Strange
  molecular partners of the $Z_c$(3900) and $Z_c$(4020)}},
  \href{https://doi.org/10.1103/PhysRevD.103.074029}{\emph{Phys. Rev. D}
  {\bfseries 103} (2021) 074029}
  [\href{https://arxiv.org/abs/2011.08725}{{\ttfamily 2011.08725}}].

\bibitem{Baru:2021ddn}
V.~Baru, E.~Epelbaum, A.A.~Filin, C.~Hanhart and A.V.~Nefediev, \emph{{Is
  $Z_{cs}(3982)$ a molecular partner of $Z_{c}(3900)$ and $Z_{c}(4020)$
  states?}}, \href{https://doi.org/10.1103/PhysRevD.105.034014}{\emph{Phys.
  Rev. D} {\bfseries 105} (2022) 034014}
  [\href{https://arxiv.org/abs/2110.00398}{{\ttfamily 2110.00398}}].

\bibitem{Ortega:2021enc}
P.G.~Ortega, D.R.~Entem and F.~Fernandez, \emph{The strange partner of the
  {{Z{\textsubscript{c}}}} structures in a coupled-channels model},
  \href{https://doi.org/10.1016/j.physletb.2021.136382}{\emph{Phys. Lett. B}
  {\bfseries 818} (2021) 136382}
  [\href{https://arxiv.org/abs/2103.07871}{{\ttfamily 2103.07871}}].

\bibitem{Du:2022jjv}
M.-L.~Du, M.~Albaladejo, F.-K.~Guo and J.~Nieves, \emph{Combined analysis of
  the {{Z}}{\textsubscript{c}}(3900) and the {{Z}}{\textsubscript{cs}}(3985)
  exotic states},
  \href{https://doi.org/10.1103/PhysRevD.105.074018}{\emph{Phys. Rev. D}
  {\bfseries 105} (2022) 074018}
  [\href{https://arxiv.org/abs/2201.08253}{{\ttfamily 2201.08253}}].

\bibitem{Nakamura:2021bvs}
S.X.~Nakamura, \emph{{X structures in $B^+\to J/\psi\phi K^+$ as one-loop and
  double-triangle threshold cusps}},
  \href{https://doi.org/10.1016/j.physletb.2022.137486}{\emph{Phys. Lett. B}
  {\bfseries 834} (2022) 137486}
  [\href{https://arxiv.org/abs/2111.05115}{{\ttfamily 2111.05115}}].

\bibitem{Luo:2022xjx}
X.~Luo and S.X.~Nakamura, \emph{{$X$ and $Z_{cs}$ in $B^+\to J/\psi\phi K^+$ as
  $s$-wave threshold cusps and alternative spin-parity assignments to $X(4274)$
  and $X(4500)$}},
  \href{https://doi.org/10.1103/PhysRevD.107.L011504}{\emph{Phys. Rev. D}
  {\bfseries 107} (2023) L011504}
  [\href{https://arxiv.org/abs/2207.12875}{{\ttfamily 2207.12875}}].

\bibitem{Maiani:2021tri}
L.~Maiani, A.D.~Polosa and V.~Riquer, \emph{{The new resonances $Z_{cs}(3985)$
  and $Z_{cs}(4003)$ (almost) fill two tetraquark nonets of broken
  SU(3){$_f$}}}, \href{https://doi.org/10.1016/j.scib.2021.04.040}{\emph{Sci.
  Bull.} {\bfseries 66} (2021) 1616}
  [\href{https://arxiv.org/abs/2103.08331}{{\ttfamily 2103.08331}}].

\end{thebibliography}\endgroup

\end{document}